\documentclass [a4paper,twocolumn]{article}
\usepackage[cp1251]{inputenc}
\usepackage[english]{babel}
\usepackage[final]{graphicx}
\newcommand{\beq}[1]{\begin{equation} \label{#1} }
\newcommand{\eeq}   {\end{equation}}
\newcommand{\ds}{\displaystyle \mathstrut}
\newcommand{\Frac}[2]{\frac
{\textstyle\lefteqn{\phantom{{}_{\mathstrut}^{\mathstrut}}} #1}
{\textstyle\lefteqn{\phantom{{}_{\mathstrut}^{\mathstrut}}} #2}}
\newcommand{\efrac}[2]{\frac{\scriptstyle \mathstrut #1}
{\scriptstyle \mathstrut #2}}
\newcommand{\kp}{\mbox{\ae}}
\newcommand{\ve}{\varepsilon}
\newcommand{\vrho}{\mbox{\boldmath $\rho$}}
\newcommand{\VSigma}{\mbox{\boldmath $\Sigma$}}
\newcommand{\av}[1]{\langle #1 \rangle}
\begin{document}
\title{Bulk effects in the coherent inelastic scattering of ultracold neutrons}
\author{A.L.Barabanov, S.T.Belyaev\\
{\it Kurchatov Institute, 123182 Moscow, Russia}}
\date{}
\maketitle

\abstract{
With the use of theory developed earlier, bulk effects in ultracold neutron coherent
inelastic scattering are considered both for solid and liquid target samples related to energy and momentum exchange with phonon and diffusion-like modes. For the neutron in a material trap, differential and
integral probabilities for the energy transfer per bounce are
presented in a simple analytic form which exhibits  the parameter
dependence. As an example, the theoretical values for the ultracold
neutron loss rate from a storage bottle with Fomblin coated walls and stainless steel walls are evaluated. Possible contribution from incoherent inelastic scattering on hydrogen contamination is discussed.}
\bigskip

\bigskip


\section{Introduction}
In the last few decades since the discovery \cite{Lus69,Ste69}
of ultracold neutrons (UCN), great progress has been made in this field of physics.
Now one considers UCN not only as a tool for studying fundamental properties of neutrons (life time and electric dipole moment) but also as a method for investigating material surfaces and thin films \cite{Gol96,Kor04}.

The main specific feature of UCN is their repulsion from material samples,
which allows storing them for a long time in material vessels. This repulsion
results from rescattering of the neutron wave in the matter. Indeed, each atom
in the target sample feels not only the incoming neutron wave but also the
secondary waves from the other atoms. Effective secondary waves come mainly from
surrounding volume $\sim\lambda^3$, where all $\sim n\lambda^3$ scatterers add
coherently onto amplitude $\sim n\lambda^3(b/\lambda)$ ($\lambda$ is the neutron
wave length, $b$ is the neutron-nucleus scattering amplitude, and $n$ is
the density of scatterers). This amplitude may be neglected, as compared
to the incoming neutron wave, if $nb\lambda^2\ll 1$. This is just the cases
for thermal and cold neutrons but not true for ultracold neutrons.
Thus, for the ultracold neutrons rescattering in the matter is of crucial importance,
and for the wave vector $k<\sqrt{4\pi nb}$ (if $b>0$) the rescattering becomes
the dominant process and results in the total reflection from the surface.
This dominant elastic scattering is simply described by a mirror potential
$U=2\pi\hbar^2nb/m$, where $m$ is the neutron mass.

Inelastic scattering of UCN on the trap walls also takes place and,
in particular, (along with $\beta$-decay and radiative capture)
results in UCN losses from the material vessels. For a long time attention
was focused on the inelastic transition of UCN to the thermal
energy region (up-scattering), i.e. on large energy transfers (with
probabilities per bounce 10$^{-6}$ and higher). In the last few years
the inelastic scattering with small energy transfers (''small heating''
and ''small cooling'') was observed with probabilities per bounce
10$^{-8}$--10$^{-6}$ (see, e.g. \cite{Str00,Bon02,Lyc02,Ste02,Ser03}). At
the moment the data obtained for Fomblin oil as the sample appears to agree
with theory. It was shown in \cite{Lam02}, the small heating influences
substantially the UCN losses from the traps with walls coated by
Fomblin oil. The data on solids like stainless steel, beryllium and
copper are controversial, in particular, no small heating was observed
in \cite{Ser03} for non-magnetic solid substances (stainless steel, Be, Cu)
with the upper bound of $\sim 10^{-8}$ for the probability per bounce.

Therefore, the inelastic UCN scattering is now of great interest, first due to the storage of UCN in material traps and, the second, for possible spectroscopic applications, e.g. for investigation of low frequency modes in surface layers and thin films \cite{Kor04}. On the other side, theoretical tools for the calculation of the inelastic scattering of UCN are rather undeveloped. One usually starts from Van Hove formula \cite{Hove54} for the differential cross section of neutron inelastic scattering on an ensemble of target nuclei,
\beq{2.3}
\frac{d^2\sigma}{d\omega d\Omega}=
\frac{k'}{2\pi k}
\sum_{\nu,\nu'}b^*_{\nu}b_{\nu'}
\int\limits_{-\infty}^{+\infty}dt\, e^{i\omega t}
\av{i|e^{-i{\bf Q}\hat{\bf R}_{\nu}(t)}
e^{i{\bf Q}\hat{\bf R}_{\nu'}(0)}|i}.
\eeq
Here and below we assume that ${\bf k}$ ($\ve=\hbar^2k^2/2m$) and ${\bf k}'$
($\ve'=\hbar^2k'^2/2m$) are the wave vectors (energies) for
incident and scattered neutrons, respectively,
$\hbar\omega=\ve-\ve'$ and ${\bf Q}={\bf k}-{\bf k}'$ are the
neutron energy and momentum transfers, $\hat{\bf R}_{\nu}(t)$ is
the time dependent Heisenberg operator of the $\nu$-th nucleus
position vector, and $b_{\nu}$ is the scattering amplitude on $\nu$-th nucleus.
The averaging is understood over the initial state $|i\rangle$ of the nuclear ensemble.

Note, however, that the Van Hove formula (\ref{2.3}) is based on the Born approximation (see, e.g. \cite{Gur68,Lov84}) and, hence, does not imply rescattering. Strictly speaking, it may be applied only to thermal and cold neutrons (see also \cite{Bar00}).

Nevertheless, it is being used for UCN as the starting point for some tricky ansatz. First, one notes that the inelastic cross section, both for coherent and incoherent scattering, predicted by (\ref{2.3}) is proportional to the total number of nuclei in the target. This allows to introduce the inelastic cross section for one nucleus. It should be stressed, that this definition is pure formal, because in fact the neutron is scattered by the whole sample as directly indicated by the presence of the correlation function in the right-hand side of (\ref{2.3}). Then the resulting cross section for neutron-nucleus scattering may be somehow extrapolated from cold to ultra-cold energy region (say assuming $1/v$ law). The sum of this inelastic cross section and that for the radiative capture on one nucleus is considered as the cross section of the total loss, which is supposed to define the imaginary part of the neutron-matter optical potential (see e.g. \cite{Ign90,Gol91}). Some model modification of the Van Hove theory for UCN was attempted in \cite{Blo77}.

Note, that the observed UCN losses are, as a rule, higher than that theoretically expected. With this respect one uses the term ''UCN anomalous losses'' (see, e.g. \cite{Ign00,Ser05,Bar05}). In fact, while one speaks of a discrepancy between experiment and theory, there is really a lack of a reliable theoretical description for UCN losses. We would like to emphasize, that the rescattering results in a drastic change in elastic neutron scattering when one goes from the cold to the ultra cold region. Thus, one can expect an important influence of this transition on the inelastic scattering. Thus, a theory of the inelastic scattering of UCN, based on solid grounds is, certainly, needed.

Recently, the authors presented a general theory of neutron scattering
\cite{Bar00}, valid for the whole domain of slow neutrons from thermal to
ultracold. The target sample is considered as a dynamical system with the
eigenstates $|j\rangle$ and the eigenvalues $\varepsilon_j.$ The theory is
in fact a method of the solution of the $N+1$ body Schroedinger equation for
a "neutron + target nuclei" system. The only approximation, which was used,
is based on the fact that the neutron-nucleus potential is short-range (as
compared to an interatomic distance and a neutron wave length) and deep (as
compared to neutron and target nuclei energies).

In this theory the amplitude of the
neutron wave, $\phi$, due to rescattering, is different at each nucleus,
$\nu $, of the sample and depends on the energy already transferred, i.e. on
the target sample state, $j$. In other words, the quantity $\phi^j_{\nu}$ is
the neutron amplitude on the surface of the $\nu$-th nucleus provided that
the target sample is at the state $j$. These amplitudes are determined by
a set of linear equations. The differential cross section takes the form:
\beq{2.1}
\begin{array}{l}
\Frac{d^2\sigma}{d\omega d\Omega}= \Frac{k'}{2\pi k}
{\ds\sum_{\nu,\nu'}\sum_{j,j'}}\phi^{j*}_{\nu}\phi^{j'}_{\nu'}
\times\\[\bigskipamount]
\phantom{} \times {\ds\int\limits_{-\infty}^{+\infty}}dt\,
e^{i\omega t+i(\ve_i-\ve_j)t/\hbar} \av{j|e^{-i{\bf Q}\hat{\bf
R}_{\nu}(t)} e^{i{\bf Q}\hat{\bf R}_{\nu'}(0)}|j'}.
\end{array}
\eeq

It was shown that for thermal and cold neutrons, when rescattering
is not important,
the equations for $\phi^j_{\nu}$ become very simple and result in
$\phi^j_{\nu}\simeq\delta_{ij}b_{\nu}$.
Thus, the cross section takes the Van Hove form (\ref{2.3}).
For the opposite case of ultracold neutrons, when rescattering becomes
the dominant process, we have solved the equations for the amplitudes
$\phi^j_{\nu}$. These amplitudes were presented as expansions in a small parameter
${\bf Q} {\bf u}$, where ${\bf u}_{\nu}$ is the shift from the
equilibrium position $\vrho_{\nu}$ of the $\nu$-th nucleus in
the target sample (${\bf R}_{\nu}=\vrho_{\nu}+{\bf u}_{\nu}$).

The zero order in ${\bf Q} {\bf u}$ corresponds to the scattering
on the target sample with fixed, frozen (or infinitely heavy) nuclei and
evidently can describe only elastic scattering. For
inelastic scattering, one should consider the next orders in the small
parameter ${\bf Q} {\bf u}$. It is convenient to introduce the
renormalized neutron amplitudes
$\psi_j(\nu)\equiv
e^{i{\bf k}\vrho_{\nu}}\phi^j_{\nu}/\beta_{\nu}$,
and present them as series in ${\bf Q} {\bf u}$
\beq{2.6}
\psi_j(\nu)=\delta_{ij}\psi(\nu)+\psi_j^{(1)}(\nu)+\ldots
\eeq
Here $\beta_{\nu}$ is the scattering length on the bound nucleus
\cite{Bar00} (this quantity slightly differs from $b_{\nu}$).

Zero order amplitudes $\psi(\nu)=\psi(\vrho_{\nu})$ in the continuous
media approximation were found to satisfy the Schroedinger equation
with the optical potential $U(\vrho)=2\pi\hbar^2n(\vrho)\beta_c/m$,
where $\beta_c$ is the average ("coherent") scattering amplitude.
Thus, the solution $\psi({\bf k},\vrho_{\nu})$ determines the zero
order neutron amplitude on the surface of the $\nu$-th nucleus
inside a sample provided
that the incident neutron has the wave vector ${\bf k}$.

Dynamical properties of the sample, and therefore the inelastic
scattering, arise in the first order in ${\bf Q} {\bf u}$. The corresponding cross
section was obtained in the form
\beq{2.9}
\begin{array}{l}
\Frac{d^2\sigma_{ie}}{d\omega d\Omega}\sim
{\ds\sum_{\nu,\nu'}}\,\beta^*_{\nu}\beta_{\nu'} \nabla^i_{\nu}
\left(\bar\psi({\bf k}',\nu)\psi({\bf k},\nu)\right)^* \times{}
\\[\bigskipamount]
\phantom{} \times \nabla^j_{\nu'} \left(\bar\psi({\bf
k}',\nu')\psi({\bf k},\nu')\right)
{\ds\int\limits_{-\infty}^{+\infty}} \av{i|\hat u^i_{\nu}(t) \hat
u^j_{\nu'}(0)|i}e^{i\omega t}dt,
\end{array}
\eeq
where $\bar\psi({\bf k},\nu)=\psi(-{\bf k},\nu)$ is to be
interpreted as the wave function in the exit channel.
Needless to say that the cross section (\ref{2.9}) should
be averaged over target initial states $|i\rangle$ with thermal
equilibrium density matrix. Below such averaging is
implied. We emphasize that the approximation ${\bf Q} {\bf u}\ll 1$, which
was used to derive Eq. (\ref{2.9}), is valid for a wide range
of neutron energies, from the ultra cold to the thermal region. Thus,
all transitions within this range can be investigated on the same base.

In this paper we consider the case when the energies of initial neutrons are less than the barrier energy $U$. Therefore, the function $\psi({\bf k},\vrho)$ decreases rapidly inward from the surface of the target, namely at the lenght on the scale of 10 nm. Then, clearly, the notions like the total number of nuclei-scatterers and the inelastic cross section for one nucleus are senseless. We are not introducing them. In this sense, our approach is quite different from that commonly used for UCN \cite{Ign90, Gol91}.

We are not introducing and discussing the imaginary part of the optical potential, caused by the inelastic scattering. In our approach \cite{Bar00}, the functions of zero approximation, $\psi({\bf k},\nu)$ and
$\psi({\bf k}',\nu)$, are determined by optical potential with the imaginary part, defined only by radiative capture. At the same time, Eq.(\ref{2.9}) gives the cross section of inelastic neutron scattering on the whole target, even on macroscopic one.

Notice that in the paper \cite{Bar00} we have performed straightforward calculation of the total elastic cross section for a macroscopic plane sample and subbarier neutrons with the use of Eq.(\ref{2.1}). It was shown that it coincides with the cross section of the sample seen by incoming neutrons (as it should be). Thus, the probability of inelastic scattering per one bounce by macroscopic sample can be calculated as the ratio of the inelastic and elastic cross sections. Just in this way the corresponding probability is calculated in our approach.

However, experiments with UCN in material traps do not
allow a direct
measurement of the differential cross section (\ref{2.9}) or
differential probability per bounce. Parameters which can be
measured are the total loss rate (in particular, into the thermal region)
and the energy distributions of pre- and post-storing neutrons.
Since the probability of the small energy transfer is small, the
multiple inelastic scattering of the same neutron may be neglected.
Therefore, the energy spectrum of the stored neutrons can be
directly linked to the total number of bounces with given material
during the storage as well as to
the probability of energy transfer per one
bounce. In Section \ref{UCN storage in material trap}
we introduce the quantities measured in experiments,
which will be analised and estimated.

The purpose of this paper is to demonstrate a real
possibility to use our approach
for analysis of the results of specific experiments on inelastic scattering of UCN. As the
first step we assume that at the intrusion length of
$\sim 10$~nm bulk properties of the matter are of dominant importance
(discussion of the surface effects in the inelastic scattering,
e.g. contribution from visco-elastic waves for Fomblin oil or
from hydrogen thin films, see in \cite{Kor04,Lam02,Pok99a,Pok99b}).
Thus, in a uniform material the diagonal matrix element
$\av{i|\hat u^i_{\nu}(t)\hat u^j_{\nu'}(0)|i}$ may exhibit
a spatial dependence only as a function of
$\vrho_{\nu}-\vrho_{\nu'}$ and therefore allows a Fourier transform
\beq{2.10}
\av{i|\hat u^i_{\nu}(t)\hat u^j_{\nu'}(0)|i}=
\sum_{{\bf q},\omega}e^{i{\bf q}
(\vrho_{\nu}-\vrho_{\nu'})-
i\omega t}\Omega^{ij}({\bf q},\omega),
\eeq
where $\sum_{{\bf q}}=\int d^3q/(2\pi)^3$ and
$\sum_{\omega}=\int d\omega/2\pi$. The correlation
function $\Omega^{ij}({\bf q},\omega)$ for the target material
is the most uncertain factor in our approach.
We analize it in Section \ref{Correlation function}.

Besides the usual phonon (sound) modes, we consider in details slow diffusion-like modes, namely, the contribution from thermo-diffusion and low frequency transverse modes in liquids. The idea is evident. Just the slow modes may be responsible for small energy transfer. However, for the sake of completeness, we perform calculation for all possible final neutron energies, from zero up to the thermal region. Earlier, the role of the thermo-diffusion in transition to the thermal region in solids was estimated in \cite{Blo77} and found insignificant.

Really, for crystals, the dominating contribution to the up-scattering (into the thermal region) is due to phonon processes with the momentum transfer to the lattice. Our result for this contribution coincides with that obtained early in a model approach, quite different from ours (see e.g. \cite{Ign00}).

All other processes, phonon- and thermo-diffusion ones, without lattice involvement, give small contribution into the scattering with large energy transfer. On the other hand, just these non-lattice processes are responsible for small energy transfers. In this paper, the corresponding probabilities for solids are calculated for the first time.

Moreover, in liquids and amorphous matter, non-lattice processes are the only ones responsible for the energy transfers, both small and large. Thus, in this paper, the contribution of bulk modes both to small and large energy transfers for liquids and amorphous samples are estimated also for the first time.

In the expression (\ref{2.9}) we take into account all
terms with $\nu\ne\nu'$ (as well as with $\nu=\nu'$), i.e. we
calculate the coherent contribution to inelastic scattering. Thus,
we use the average value (coherent scattering length) $\beta_c$
instead $\beta_{\nu}$ in expressions for the inelastic cross section
derived from (\ref{2.9}).  Note, that
to evaluate the incoherent contribution related both to spin-flip and
mixture of nuclei with different scattering lengths, one needs to
separate additional term (proportional to $\beta_{inc}^2$) from the
general equation (\ref{2.9}) keeping only $\nu=\nu'$.

In this paper we restrict ourselves to the coherent inelastic scattering.
Thus, we realize that our consideration is unsuitable for treatment of
neutron scattering on targets with high hydrogen contamination. Our main
goal is application to "hydrogen-free" materials.

As the starting point we use the following expression for the coherent inelastic cross section \cite{Bar00}:
\beq{2.11}
\frac{d\sigma_{ie}}{d^3k'}=
\frac{\hbar}{2\pi mk}
\sum_{{\bf q}}B^{i*}({\bf q})B^j({\bf q})
\Omega^{ij}({\bf q},\omega),
\eeq
where
\beq{2.12}
{\bf B}({\bf q})=\beta_c
\sum_{\nu} e^{-i{\bf q}\vrho_{\nu}}
{\bf\nabla}_{\nu}\left(\bar\psi({\bf k}',\nu)
\psi({\bf k},\nu)\right).
\eeq
It describes the transition of an incident neutron with the wave vector ${\bf k}$ into an
element $d^3k'$ of the final wave vector space ${\bf k}'$.
This equation is valid for neutrons in a broad energy region,
since the smallness of the parameter ${\bf Q}{\bf u}$, that was
used for deriving Eq.(\ref{2.11}), is valid even for thermal
neutrons. Indeed, for the latter, the scattering on the
optical potential ($\sim 100$~neV) is not important. Then, if one
replaces the functions $\psi({\bf k},\vrho)$ and $\bar\psi({\bf
k}',\vrho)$ by the plane waves $e^{i{\bf k}\vrho}$ and
$e^{-i\bar{\bf k}'\vrho}$, respectively, then (\ref{2.11})
transforms onto (\ref{2.3}).

In this paper we consider
the inelastic scattering of initial ultracold neutrons
to any final energies, below and above the potential barrier.
Of course, to calculate the neutron inelastic cross section
from (\ref{2.11}) and (\ref{2.12}) one needs a specific model
for the target sample to find solutions  $\psi({\bf k},\vrho)$
and $\bar\psi({\bf k}',\vrho)$ for input and output channels.
This problem is considered in Section \ref{Specific target sample}.

In section \ref{Inelastic cross section} it is shown that multi-dimensional integration in (\ref{2.11}) and (\ref{2.12}) can be performed explicitly with realistic correlation functions. The details of the calculation are presented in Appendix. In section \ref{s7} we analyze how analytical results for differential probability are changed when specified for the excitation mode and energy transfer. In section \ref{s8} numerical results are presented for two typical substances, stainless steel and Fomblin oil. Results obtained, both analytical and numerical, are discussed in section \ref{Discussion}. Conclusion is given in section \ref{s9}.

\section{UCN storage in material trap
\label{UCN storage in material trap}}

\subsection{Probability for energy transfer per one bounce}

The probability for energy transfer per one bounce is naturally
defined as a ratio of two cross sections, differential for energy
transfer and the integral elastic one. The latter, in our case, is
equal simply to the total sample area seen by the neutron,
$S_{\perp}=Sk_{\perp}/k=S\cos\theta$, $\theta$ is the angle of
incidence. With fixed initial wave vector ${\bf k}$ (or two its
components $k_{\perp}$ and $k_{\|}$), the probability for the
neutron after a bounce to have the kinetic energy $\varepsilon'$ is
\beq{120a}
\frac{dw(k_{\perp},k_{\|}\to\varepsilon')}{d\varepsilon'}=
\frac{1}{S_{\perp}}\int
\frac{d\sigma_{ie}}{d^3k'}\,\frac{d^3k'}{d\varepsilon'}.
\eeq

In reality, the inelastic scattering of UCN with fixed ${\bf k}$
cannot be observed. One usually measures quantities somehow
averaged over the initial momentum ${\bf k}$. Let us introduce a
quasi-classical distribution $F({\bf r},{\bf k})$ in position
${\bf r}$ and momentum ${\bf k}$ of UCN inside the trap,
normalized by the condition \beq{7n.6} \int d^3r\int d^3k\,F({\bf
r},{\bf k})=N_0, \eeq where $N_0$ is the total number of UCN in
the trap.

The number of neutrons, scattered by the element
$d{\bf S}$ of the material sample
at the position ${\bf r}_S$ (and height $h_S$)
to the interval $dE'$ of the total energy $E'=\ve'+mgh_S$ during the
time $dt$, is given by
\beq{130a}
\begin{array}{l}
dN=dE'dt{\ds\int} d^3k\, F({\bf r}_S,{\bf k})\,
\left({\bf v}\,d{\bf S}\right)\,\times{}
\\[\bigskipamount]
\phantom{dN=dE'dt}\times
\Frac{dw(k_{\perp},k_{\|}\to E'-mgh_S)}{dE'}\,,
\end{array}
\eeq
where ${\bf v}=\hbar {\bf k}/m$ is the neutron velocity, and $mgh_S$
is the neutron potential energy in the earth's gravitational field.
The total rate
of inelastic transition to the energy interval $dE'$ is of the
form
\beq{7n.9}
\begin{array}{l}
\Frac{dN(E')}{dE'dt}= {\ds\oint}dS
{\ds\int\limits_{k_{\perp}>\,0}}
d^3k\, v_{\perp}F({\bf r}_S,{\bf k})\times
\\[\bigskipamount]
\phantom{\Frac{dN(E')}{dE'dt}= {\ds\oint}dS}\times
\Frac{dw(k_{\perp},k_{\|}\to E'-mgh_S)}{dE'}\,.
\end{array}
\eeq

Let us assume that the momentum distribution is iso\-tro\-pic, i.e.
$F({\bf r},{\bf k})=F({\bf r},k)$. Then, it is convenient to
introduce a distribution $f({\bf r},\ve)$ in position and kinetic
energy of UCN inside the trap
\beq{7n.10}
f({\bf r},\ve)=
2\pi\left(\frac{2m}{\hbar^2}\right)^{3/2}\sqrt{\ve}\,
F({\bf r},k),\quad
k=\sqrt{\frac{2m\ve}{\hbar^2}},
\eeq
normalized by the condition
\beq{7n.11}
\int d^3r\int\limits_0^{\infty}d\ve
f({\bf r},\ve)=N_0.
\eeq
Then, the transition rate takes the form
\beq{7n.12}
\frac{dN(E')}{dE'dt}= \oint
dS\int\limits_0^{\infty}d\ve\, vf({\bf r}_S,\ve)
\frac{dw(\ve\to E'-mgh_S)}{dE'},
\eeq
where
\beq{7n.13}
\frac{dw(\ve\to\ve')}{d\ve'}=
\av{\frac{dw(k_{\perp},k_{\|}\to\ve')}{d\ve'}}
\eeq
is the differential probability of inelastic scattering averaged
over the angle of incidence $\theta$.

Here and below averaging of any function
$f(k_{\perp},k_{\|})$ over $\theta$ is defined as
\beq{7n.13.2}
\av{f(k_{\perp},k_{\|})}=
\frac{1}{2}\int\limits_0^{\pi/2}
d\theta\sin\theta\cos\theta\,
f(k\cos\theta,k\sin\theta).
\eeq

In the simplest approximation we neglect the earth's gravitational
field, and assume, first, the density $n_U$ of UCN in the trap is
uniform over the volume and, second, the energy of UCN is fixed and
equal $\ve_U$,
thus
\beq{7n.16}
f({\bf r},\ve)=n_U\delta(\ve-\ve_U).
\eeq
The inelastic transition rate is of the form
\beq{7n.17}
\frac{dN(\ve')}{d\ve'dt}=
Sn_Uv_U \frac{dw(\ve_U\to\ve')}{d\ve'}
\eeq
where $S$ is the total area of the material surface inside the trap.

\subsection{Probability for neutron losses from a trap}

When the final neutron energy $E'$ exceeds the barrier
energy $U$, the neutron escapes from the trap. The probability of
this event and the corresponding transition rate can be obtained as
an integral over $E'$ from $U$ up to infinity from
(\ref{7n.12}) (in the simplest approximation -- as the similar
integral over $\ve'$ from (\ref{7n.17})).

The total escape probability is one of the quantities of practical
interest. The second quantity of our interest is the small energy
transfer. To be compared with measured values, both probabilities
should be properly averaged over the energy and space distributions of
the stored neutrons, and corrected for escape during the storage. Such
averaging depends on the details of any specific experiment. On
the other hand, the simple estimate (\ref{7n.17}) for
the inelastic transition rate
and its integral over $\varepsilon'$ allow useful rough comparison
with measured values.

\section{Correlation function
\label{Correlation function}}

In this section we consider the correlation function
$\Omega^{ij}({\bf q},\omega)$, entering Eq.(\ref{2.11}).
Correlation functions describe response (relaxation) of a substance
after distortion of its statistical equilibrium by some external
force characterized by $\omega$ and ${\bf q}$. Their structures
for solids and liquids are covered in many books and articles on
fluctuations and kinetics (see, e.g. \cite{Sch68,For75,Akh81,Lan86}).
We restrict ourselves to the details necessary for what follows.

Fast distortions (when $|\omega|\gg\tau^{-1}$, where $\tau$ is
some average relaxation time) are relaxed by phonons (or sound
waves). In this region the correlation function can be easily
obtained by expansion of the displacement vectors $\hat{\bf
u}^i_{\nu}(t)$ in phonon amplitudes. The result is well known
(see, e.g. Eq. (4.26) of Ref. \cite{For75}):
\beq{3n.1}
\Omega^{ij}({\bf q},\omega)=
\Omega_l^{ij}({\bf q},\omega)+
\Omega_t^{ij}({\bf q},\omega),
\eeq
where the
longitudinal and transverse parts are given by
\beq{3n.2}
\Omega_l^{ij}({\bf q},\omega)=
\frac{q_iq_j}{q^2}\Omega_l(q,\omega),
\eeq
\beq{3n.2.2}
\Omega_t^{ij}({\bf q},\omega)=
\left(\delta_{ij}-\frac{q_iq_j}{q^2}\right)
\Omega_t(q,\omega),
\eeq
and
\beq{3n.3}
\Omega_{l,t}(q,\omega)=
\frac{2\hbar n_0(\omega)}{\rho c^2_{l,t}}\,
\frac{\Gamma^2_{l,t}}
{(q^2-\omega^2/c^2_{l,t})^2+\Gamma^4_{l,t}}.
\eeq
Here $\rho=Mn$ is the material density ($M$ is the nuclear
mass), and
\beq{3n.4}
n_0(\omega)=
\frac{1}{\exp (\hbar|\omega|/T)-1}\quad
\mathrel{\mathop{\simeq}\limits_{\hbar|\omega|\ll T}}\quad
\frac{T}{\hbar|\omega|}
\eeq
is the occupation factor for a mode
with the frequency $\omega$ at the temperature $T$. We allow the
longitudinal (l) and transverse (t) sound velocities $c_{l,t}$ and
damping factors $\Gamma_{l,t}$ to be different.

For isotropic solids, the factors $\Gamma^2_{l,t}$ can be taken
propor\-ti\-onal to the absorption coefficients of longitudinal and
transverse sound (see, e.g. \cite{Lan86b}) \beq{3n.5}
\Gamma^2_l=\frac{\gamma |\omega|^3}{c^4_l},\quad
\gamma=\frac{\zeta}{\rho}+\alpha D_S, \eeq \beq{3n.6}
\Gamma^2_t=\frac{\nu |\omega|^3}{c^4_t},\quad
\nu=\frac{\eta}{\rho}. \eeq Here $\eta$ and
$\zeta=4\eta/3+\varphi$ are the dynamic shear and longitudinal
viscosities ($\varphi$ is the volume viscosity), $\nu$ is the
kinematic viscosity, $\alpha=C_P/C_V-1$, where $C_P$ and $C_V$ are
the specific heats for constant pressure and volume, respectively,
$D_S=\kp/\rho C_P$ is the thermo-diffusion coefficient ($\kp$ is
the thermo-conductivity).

To describe a small energy exchange between UCN and a sample, the form
of correlation function for small $\omega$ and $q$ is of prime
interest. Slow and long-range fluctuations are related to coherent
motion of a great number of media atoms (hydrodynamic modes). For
such kinds of distortion, local statistical equilibrium comes first
and  macro-relaxation by hydrodynamic processes follows.
There\-fo\-re, in the region of small $\omega $ and $q$ correlation
functions for solids and liquids are governed by hydrodynamics and
should depend on its parameters. Two different approaches, namely,
phenomenological (from hydrodynamic fluctuations and Kubo
theory \cite{Sch68,For75}) and quantum many-body theory
\cite{Bal69}, give the same result. In hydrodynamics two kind of
processes (and corresponding correlations) are distinguished,
longitudinal and transverse.

The longitudinal correlation function is related to
"den\-si\-ty--density" fluctuations \beq{3n.7} \frac{1}{2\pi
N}\av{\int \hat n({\bf r}'+{\bf r},t)\,\hat n({\bf r}',0)\,d{\bf
r}'\,}= \sum_{{\bf q},\omega}e^{i{\bf q}{\bf r}-i\omega t} S({\bf
q},\omega), \eeq where a Fourier transform, the "dynamical structure factor",
can be shown to be connected with $\Omega_l({\bf q},\omega)$ by
\beq{3n.8} S({\bf q},\omega)\simeq\frac{nq^2}{2\pi} \Omega_l({\bf
q}',\omega). \eeq The vector ${\bf q}'$ in $\Omega_l$ is in the
first Brillouin zone and ${\bf q}$, the momentum transfer, may
differ from ${\bf q}'$ by a reciprocal lattice vector. In the
right-hand side of (\ref{3n.8}), we omit all terms proportional to
$\delta(\omega)$ and related to elastic scattering.

Following \cite{Sch68,For75} we get both for solids and liquids
\beq{3n.9}
\Omega_l(q,\omega)= \frac{T}{i\rho\omega}
\left(\frac{1}{\omega^2-c^2_{lT}q^2-i\omega\Gamma_l(q,\omega)}-
{\rm c.c.}\right),
\eeq
which looks like a phonon-type but with
complicated form of "phonon absorption"
\beq{3n.10}
\Gamma_l(q,\omega)=
\frac{(c^2_{lS}-c^2_{lT})q^2}
{D_Tq^2+i\omega}+
\frac{\zeta q^2}{\rho},
\eeq
where $D_T$ is the thermo-diffusion coefficient.
The subscripts $T$ and $S$ indicate constant temperature and
entropy, respectively, and \beq{3n.11} \frac{c^2_{lS}}{c^2_{lT}}=
\frac{D_T}{D_S}= \frac{C_P}{C_V}= 1+\alpha. \eeq

For $q^2\ll |\omega|/D_T$, Eqs. (\ref{3n.9}) and (\ref{3n.10})
result in the pure phonon-type correlation function with the sound
velocity $c_{lS}$ and the absorption defined by (\ref{3n.5}). For
arbitrary small $\omega$ and $q$ (hydrodynamic region)
\beq{3n.12}
|\omega|\ll\frac{c_l^2}{\gamma},\quad
q\ll\frac{c_l}{\gamma},
\eeq
the function (\ref{3n.9}) takes the form
\beq{3n.13}
\begin{array}{l}
\Omega_l(q,\omega)=
\Frac{iT}{\rho\omega}\times{}
\\[\bigskipamount]
\phantom{}\times \left(\Frac{q^2+i\omega/D_T}
{c^2_{lT}(q^2+i\omega/D_S)
(q^2-\omega^2/c^2_{lS}+i\omega^3\gamma/c^4_{lS})}-
{\rm c.c.}\!\right).
\end{array}
\eeq
It has two kind of poles, of
phonon ($p$) and of thermo-diffusion ($d$) origin, and it is useful to
separate them as
\beq{3n.14}
\Omega_l(q,\omega)=
\Omega_l^p(q,\omega)+ \Omega_l^d(q,\omega).
\eeq
Here $\Omega_l^p(q,\omega)$ is given by (\ref{3n.3})
with $\Gamma_l^2$ (\ref{3n.5}) and
\beq{3n.14.2}
\Omega_l^d(q,\omega)=
\frac{2T}{\rho c_{lS}^2|\omega|}\,
\frac{\alpha\Gamma_d^2}{q^4+\Gamma_d^4},\quad
\Gamma_d^2=\frac{|\omega|}{D_S}.
\eeq

It is easy to see that additional diffusion-type pole in
(\ref{3n.13}) originates from the pole in the ''phonon
absorption'' $\Gamma_l(q,\omega)$ (\ref{3n.10}). The complicated
structure of $\Gamma_l(q,\omega) $ is intrinsic only for
longitudinal phonons.

The transverse correlation function for solids has only one,
phonon-like, pole. So, transverse excitations in solids are
relaxed only by phonons, and the correlation function for small
$\omega$ and $q$ has the same form (\ref{3n.3}) as for large $q$
and $\omega$.

In common liquid models, kinetics is assumed to go by finite jumps
of atoms from one equilibrium position to the other after some
average waiting time $\tau$. For high frequencies
$|\omega|\gg\tau^{-1}$ a liquid behaves like a solid. Low frequency
waves ($|\omega|\ll\tau^{-1}$) damp due to viscosity. This limit
is reproduced by the correlation function for the liquid model
with one relaxation time (see, e.g. \cite{Sch68})
\beq{3n.15}
\Omega_t(q,\omega)=
\frac{T}{i\rho c^2_tq^2}
\left(\frac{1}{\omega-i\Gamma_t(q,\omega)}-{\rm c.c.}\right),
\eeq
where
\beq{3n.16}
\Gamma_t(q,\omega)=\frac{c^2_tq^2\tau}{1+i\omega\tau}. \eeq The
relaxation time is usually estimated as $\tau=\eta/G=\nu/c_t^2$,
where $\eta$ and $\nu=\eta /\rho$ are dynamic and kinematic shear
viscosities, respectively, $G$ is the modulus of elasticity, and
$c_t^2=G/\rho$.

The correlation function (\ref{3n.15}) and (\ref{3n.16})
exactly transforms to the form (\ref{3n.3}) with
the damping factor
\beq{3n.18}
\Gamma^2_t=\frac{|\omega|}{c^2_t\tau}=
\frac{|\omega|}{\nu}.
\eeq
Note, that for low frequencies
\beq{3n.18.2}
|\omega|\ll\tau^{-1}=\frac{c_t^2}{\nu}
\eeq
there exists a wide hydrodynamic region for $q$
\beq{3n.18.3}
\frac{|\omega|}{c_t}\ll q\ll\frac{c_t}{\nu},
\eeq
where the Eq.(\ref{3n.15}) (as well as (\ref{3n.3}))
transforms into the diffusion-type function (\ref{3n.14.2}) with
damping factor (\ref{3n.18}). It is of interest that this factor
is proportional to $\nu^{-1}$ in contrast to (\ref{3n.6}) (see, e.g.
\cite{Fre46}).

The correlation functions can be determined from first
principles only for two limiting areas of variables, large and
small $\omega$ and ${\bf q}$. In the intermediate area one should
use some specific interpolation models for the substance
considered.

Thus, we use the correlation function as a sum
of phonon- and diffusion-type terms. Contribution to the
cross-section from each part can be calculated independently and
summed afterwards.

For the calculation of the cross-section we will need $\Omega({\bf
q},\omega)$ as a function of $q_{\perp}$
($q^2=q_{\perp}^2+q_{\|}^2$), and it is convenient to present all
cases considered above in the same form
\beq{3n.19}
\Omega(q_{\perp})=
\frac{2\alpha\hbar n_0(\omega)}{\rho c^2}
\bar\Omega(q_{\perp}),
\eeq
where
\beq{3n.19.2}
\bar\Omega(q_{\perp})=
\frac{\Gamma^2}{(q^2_{\perp}+p^2)^2+\Gamma^4},\quad
p^2=q^2_{\|}-\frac{\epsilon\,\omega^2}{c^2},
\eeq
and
$\alpha=\epsilon=1$ for all cases besides the longitudinal
thermo-diffusion mode. In the latter case we have
$\alpha=C_P/C_V-1$ and $\epsilon=0$.

Note, that $\Gamma_p^2$ (phonon absorption)  is a small parameter for
all values of
$\omega$ and $q$ that we are considering, and when
$\Gamma_p\to 0$,
\beq{3n.21}
\bar\Omega_p(q_{\perp})
\quad\longrightarrow\quad
\pi\delta\left(q^2-\frac{\omega^2}{c^2}\right),
\eeq
in agreement with the phonon correlation function usually used.

\section{Specific target sample: uniform thick layer
\label{Specific target sample}}

In this section we will derive the function ${\bf B}({\bf q})$ (\ref{2.12})
  for a simple model commonly relevant to the UCN storage experiments,
i.e.  an uniform thick layer. The appropriate choice of the model
allows us to define  the solutions $\psi({\bf
k},\vrho)$ and $\bar\psi({\bf k}',\vrho)$ of the
Schroedinger equation with optical potential
for input and output channels.

The problem is that the function
$\psi({\bf k},\nu)$, being the neutron amplitude at the $\nu$-th nucleus,
is defined only inside the sample. But to get $\psi({\bf
k},\nu)$ in the continuous media approximation, one has to find the
scattering type solution of the Schroedinger equation
with an external
plane wave entering the sample, $e^{i{\bf k}\vrho}$. For
the input channel function, $\psi({\bf k},\nu)$, one may imagine the
external plane wave as attributed to the incoming neutron. Let us assume it
to come from the left. For the output channel function,
$\bar\psi({\bf k}',\vrho)=\psi(-{\bf k}',\vrho)$, the
corresponding external wave, $e^{-i{\bf k}'\vrho}$, is an
auxiliary quantity and should be directed to the right, for back
scattering, and to the left, for forward scattering.

To include both cases, let us consider a uniform layer located at
$-a/2<z<a/2$ with the surface area $S$ and the thickness $a$, much
larger than the incident neutron wave length. The sample is
supposed to have the density $n$ and the scattering length $\beta_c$
providing a constant
optical potential $U$. Note that in a general case,
$\beta_c$  is complex with an imaginary part that can be used to
describe the radiative capture of UCN.

We consider two solutions of the model scattering problem,
described above, with the left (L) and right (R) incoming waves.
Outside the target they have the form \beq{3.1.2} \psi^{(L)}({\bf
k},{\bf r})= e^{i{\bf k}_{\|}{\bf r}_{\|}}\cdot
\left\{\begin{array}{ll} e^{ik_{\perp}z}+R\,e^{-ik_{\perp}z}, &
z<-a/2,
\\
T\,e^{ik_{\perp}z}, & z>a/2,
\end{array}\right.
\eeq \beq{3.1.3} \psi^{(R)}({\bf k},{\bf r})= e^{i{\bf k}_{\|}{\bf
r}_{\|}}\cdot \left\{\begin{array}{ll} T\,e^{-ik_{\perp}z}, &
z<-a/2,
\\
e^{-ik_{\perp}z}+R\,e^{ik_{\perp}z}, & z>a/2.
\end{array}\right.
\eeq In both cases $k_{\perp}\equiv |{\bf k}_{\perp}|>0$ is the
normal component of the neutron momentum, and ${\bf k}_{\|}$ is
its component along the target surface.

The functions inside the layer, determined by the boundary
conditions, can be written (for constant U) as \beq{3.2}
\begin{array}{l}
\psi^{(L,R)}({\bf k},\nu)= e^{i{\bf k}_{\|}{\bf r}_{\|}}
e^{-ik_{\perp}a/2} \times{} \\[\bigskipamount]
\phantom{\psi^{(L,R)}({\bf k},\nu)}\times
\Frac{t\gamma^{1/2}}{1-r^2\gamma^2} {\ds\sum_{\sigma=\pm 1}}
A^{(L,R)}_{\sigma}e^{i\sigma\bar kz}.
\end{array}
\eeq
Here $\bar k$ is the normal component of neutron momentum
inside the target sample
\beq{3.3}
\bar k=\sqrt{k^2_{\perp}-k^2_0},\quad
k_0^2=4\pi\beta_c n,
\eeq
where $k_0$ is the value of the neutron wave number $k_{\perp}$ at
the barrier, $r$ and $t$ are the reflection
and transition coefficients for the target surface \beq{3.4}
t=\frac{2k_{\perp}}{k_{\perp}+\bar k},\quad r=\frac{k_{\perp}-\bar
k}{k_{\perp}+\bar k}, \eeq and the coefficients $A$ are determined
by \beq{3.5} A^{(L)}_{+1}=1,\quad A^{(L)}_{-1}=-r\gamma\quad,
A^{(R)}_{+1}=-r\gamma,\quad A^{(R)}_{-1}=1. \eeq The factor
\beq{3.6.2} \gamma=e^{i\bar ka} \eeq oscillates very rapidly as a
function of $k_{\perp}$ above the barrier ($k_{\perp}>k_0$), and
vanishes below the barrier ($k_{\perp}<k_0$).

Below we use the "left" solution (\ref{3.1.2}) for the input
channel function. Then, the output channels with the backward and
forward scattering angles correspond to the "left", L,
(\ref{3.1.2}) and "right", R, (\ref{3.1.3}) solutions,
respectively. Inside the layer these solutions can be represented
as \beq{3.7}\begin{array}{l} \bar\psi^{(L,R)}({\bf k}',\nu)=
e^{-i{\bf k}'_{\|}{\bf r}_{\|}} e^{-ik'_{\perp}a/2}
\times{}\\[\bigskipamount]
\phantom{\bar\psi^{(L,R)}({\bf k}',\nu)} \times
\Frac{t'\gamma'^{1/2}}{1-r'^2\gamma'^2} {\ds\sum_{\sigma=\pm 1}}
A'^{(L,R)}_{\sigma}e^{i\sigma\bar k'z}.\end{array} \eeq All
momenta and momentum dependent quantities
(\ref{3.3})-(\ref{3.6.2}) for the output channel are denoted by
primes: $k'_{\perp}$, $\bar k'$, $t'$, $r'$, $A'_{\sigma}$,
$\gamma'$. They are interrelated
by the same equations as that for the
input channel.

Combining (\ref{3.2}) and (\ref{3.7}) we obtain for (\ref{2.12})
\beq{3.8}
{\bf B}({\bf q})=i\beta_ctt'
\frac{e^{-i(k_{\perp}+k'_{\perp})a/2}}
{(1-r^2\gamma^2)(1-r'^2\gamma'^2)}\,
\Sigma_{\|}\VSigma.
\eeq
Here
\beq{3.8.2}
\Sigma_{\|}=\sum_{\nu_{\|}}
e^{i({\bf Q}_{\|}-{\bf q}_{\|}){\bf r}_{\|}},\quad
{\bf Q}_{\|}={\bf k}_{\|}-{\bf k}'_{\|},
\eeq
and
\beq{3.8.3}
\VSigma=\sum_{\sigma,\sigma'}
A_{\sigma}A'_{\sigma'}
\left({\bf Q}_{\|}+Q^{\sigma\sigma'}_{\perp}{\bf e}_z\right)
\Sigma^{\sigma\sigma'}_{\perp},
\eeq
where
\beq{3.8.4}
Q^{\sigma\sigma'}_{\perp}=
\sigma\bar k+\sigma'\bar k',\quad
\Sigma^{\sigma\sigma'}_{\perp}=
(\gamma\gamma')^{1/2}
\sum_{\nu_{\perp}}
e^{i(Q^{\sigma\sigma'}_{\perp}-q_{\perp})z}.
\eeq

Summation over the nuclei on the plane parallel to the target
surface gives
\beq{3.9}
\Sigma_{\|}=
(2\pi)^2n_{\|}\sum_{{\bf G}_{\|}}
\delta^{(2)}({\bf Q}_{\|}-{\bf q}_{\|}-{\bf G}_{\|}),
\eeq
where summation over the longitudinal reciprocal
lattice vector ${\bf G}_{\|}$ arises for crystals, and $n_{\|}$ is
the number of nuclei per the unit area of the fixed plane ($z={\rm
const}$). Therefore we have the equality
\beq{3.10}
{\bf Q}_{\|}={\bf q}_{\|}+{\bf G}_{\|}
\eeq
for the longitudinal transferred momentum.

The cross section (\ref{2.11}) is proportional to
the second power of the quantity $\Sigma_{\|}$. Then,
\beq{3.10.2}
\left|\Sigma_{\|}\right|^2=
S(2\pi)^2n^2_{\|}\sum_{{\bf G}_{\|}}
\delta^{(2)}({\bf Q}_{\|}-{\bf q}_{\|}-{\bf G}_{\|}),
\eeq
where $S$ is the surface area of the sample.

For a non-crystalline substance we replace summation over nuclear
planes by integration
\beq{3.11}
\Sigma^{\sigma\sigma'}_{\perp}=
(\gamma\gamma')^{1/2}n_{\perp}
\int\limits_{-a/2}^{a/2}
e^{i(Q^{\sigma\sigma'}_{\perp}-q_{\perp})z}dz
\eeq
and obtain
\beq{3.11.2}
\Sigma^{\sigma\sigma'}_{\perp}=
(\gamma\gamma')^{1/2}n_{\perp}
\Delta(q_{\perp}-Q^{\sigma\sigma'}_{\perp}),
\eeq
where
\beq{3.12}
\Delta(x)=\frac{2\sin (xa/2)}{x}.
\eeq
Here $n_{\perp}$ is the number of nuclear planes
per the unit length along the $z$ axis ($n_{\|}n_{\perp}=n$). The
function $\Delta(x)$ peaks sharply for small value of the
argument.

For crystals the sum $\Sigma^{\sigma\sigma'}_{\perp}$ has
additional peaks when $Q^{\sigma\sigma'}_{\perp}-q_{\perp}$ is
close to the transverse reciprocal lattice vector $G_{\perp}$. In
this case we take
\beq{3.13}
\Sigma^{\sigma\sigma'}_{\perp}=
(\gamma\gamma')^{1/2}n_{\perp}
\Delta(q_{\perp}+G_{\perp}-Q^{\sigma\sigma'}_{\perp}),
\eeq
assuming the argument in $\Delta$-function is in the first
Brillouin zone. Since the vector $q_{\perp}$ (originated from
correlation function) is in the first Brillouin zone, the vector
$G_{\perp }$ should compensate, if necessary, exceeding part of
$Q^{\sigma\sigma'}_{\perp}$.

\section{Inelastic cross section and
inelastic transition probability
\label{Inelastic cross section}}

With the given correlation function (\ref{3n.19.2})
and quantity ${\bf B}$ (\ref{3.8}), the
general expression for the inelastic cross section (\ref{2.11})
is fully defined and we can easily integrate over ${\bf q}_{\|}$ with the
help of (\ref{3.10.2}).
Taking into account contributions from  both, longitudinal and transverse
parts of the correlation function, we find
\beq{5.2}
\frac{d\sigma_{ie}^{l,t}}{d^3k'}=
S \frac{\alpha n\beta_c^2\hbar^2n_0(\omega)}
{2\pi kmMc_{l,t}^2}
|t|^2|t'|^2\sum_{{\bf G}_{\|}} \Pi_{l,t}\,,
\eeq
where the
function
\beq{5.3}
\begin{array}{l}
\Pi_{l,t}=
\Frac{|\gamma|\,|\gamma'|}
{|1-r^2\gamma^2|^2\,|1-r'^2\gamma'^2|^2}\times{}
\\[\bigskipamount]
\phantom{\Pi_{l,t}}\times
{\ds\sum_{\sigma,\sigma',\tau,\tau'}}
A_{\sigma}A^*_{\tau} A'_{\sigma'}A'^*_{\tau'}\,
J_{l,t}(\lambda,\eta)

\end{array}
\eeq
includes all factors
$\gamma$ and $\gamma'$ that are strongly oscillating above the
barrier and exponentially small below it.

The factor $J_{l,t}(\lambda,\eta)$ depends on transverse momenta
inside the target in combinations
\beq{5.4}
\lambda=Q^{\sigma\sigma'}_{\perp}-G_{\lambda},\quad
\eta={Q^{\tau\tau'}_{\perp}}^*-G_{\eta},
\eeq
where $\lambda$ and $\eta$ are assumed to be in the first
Brillouin zone, and the difference from the full combinations
$Q^{\sigma\sigma'}_{\perp}$ and ${Q^{\tau\tau'}_{\perp}}^*$
(see (\ref{3.8.4})) is compensated by the transverse
components $G_{\lambda}$ and $G_{\eta}$ of the reciprocal lattice
vectors. The factor $J_{l,t}(\lambda,\eta)$ is defined by the
integral with the correlation function $\Omega_{l,t}$ and has a
different form for the longitudinal and transverse cases,
\beq{5.5}
J_{l,t}(\lambda,\eta)=
\frac{1}{\pi}\!
\int\limits_{-\infty}^{+\infty}
\!\!dq_{\perp}\bar\Omega_{l,t}(q_{\perp})
\Delta(q_{\perp}-\lambda)
\Delta(q_{\perp}-\eta)
F_{l,t}(q_{\perp}),
\eeq
where
\beq{5.5.2}
F_l(q_{\perp})=
\Frac{({\bf Q}_{\|}{\bf q}_{\|}+Q^{\sigma\sigma'}_{\perp}q_{\perp})
({\bf Q}_{\|}{\bf q}_{\|}+{Q^{\tau\tau'}_{\perp}}^*q_{\perp})}
{q_{\|}^2+q_{\perp}^2},
\eeq
\beq{5.5.3}
F_t(q_{\perp})=
Q_{\|}^2+
Q^{\sigma\sigma'}_{\perp}{Q^{\tau\tau'}_{\perp}}^*-
F_l(q_{\perp}).
\eeq
Note, that the functions $F_{l,t}(q_{\perp})$, as well as $J_{l,t}$,
depend on indices $\sigma$, $\sigma'$, $\tau$ and $\tau'$, but, to
simplify notations, we omit them.
Here and below we use ${\bf q}_{\|}$, defined from (\ref{3.10})
\beq{5.6}
{\bf q}_{\|}={\bf Q}_{\|}-{\bf G}_{\|},
\eeq
as a fixed quantity. The
correlation function $\bar\Omega(q_{\perp})$, when parameterized
as in (\ref{3n.19.2}) by $\Gamma^2$ and $p^2$, has a universal form
for the longitudinal and transverse cases, and their specificity
arises only from the numerical values of these parameters. Hence we
shall omit the subscripts on $\bar\Omega(q_{\perp})$ till the
last, numerical stage.

The inelastic transition probability (\ref{120a})
with the use of (\ref{5.2}) takes the form
\beq{7n.4}
\frac{dw_{l,t}(k_{\perp},k_{\|}\to\ve')}{d\ve'}=
\frac{\alpha\beta_ck\,n_0(\omega)}
{Mc_{l,t}^2}\,W_{l,t}(k_{\perp},k_{\|}\to\ve'),
\eeq
where the dimensionless factor $W$ is given by
\beq{7n.5}
W_{l,t}(k_{\perp},k_{\|}\to\ve')=
\frac{k_{\perp}}{\pi^2k}
\int\limits_0^{k'}dk'_{\perp}
\int\limits_0^{\pi}d\varphi\, |t'|^2\,\Pi_{l,t}.
\eeq
Here one integrates over $k'_{\perp}$ (up to
$k'=\sqrt{2m\varepsilon'}/\hbar$) and $\varphi$,
the angle between ${\bf k}'_{\|}$ and ${\bf k}_{\|}$,
with fixed $\varepsilon'$.

For crystals, the final momentum ${\bf k}'$ is split by parts
inside the first Brillouin zone, ${\bf k}'_1$, and ${\bf G}$.
So the integral (\ref{120a}) over ${\bf k}'$ should be
understood as that over ${\bf k}'_1$ and, if necessary,
a sum over ${\bf G}$.

The detailed energy distribution inside a high Brillouin zone
is of no interest, and we shall consider the full integrals
over each zone, assuming that the final energy depends only on
${\bf G}$, $\ve'\simeq\hbar^2G^2/2m$. In analogy with
(\ref{120a}), (\ref{7n.4}) and (\ref{7n.5}) we define
transition probability per bounce
\beq{8m.2}
w_{l,t}(k_{\perp},k_{\|}\to {\bf G})=
\frac{1}{S_{\perp}}\int
\frac{d\sigma^{l,t}_{ie}}{d^3k'}d^3k',
\eeq
which can be presented in the form
\beq{8m.3}
\begin{array}{l}
w_{l,t}(k_{\perp},k_{\|}\to {\bf G})=
\Frac{\alpha\beta_ck\,n_0(\hbar G^2/2m)}
{Mc^2_{l,t}}\times{}
\\[\bigskipamount]
\phantom{w_{l,t}(k_{\perp},k_{\|}\to {\bf G})}\times
\Frac{\hbar^2G^2}{2m}\,
W_{l,t}(k_{\perp},k_{\|}\to {\bf G}),
\end{array}
\eeq
where
\beq{8m.4}
W_{l,t}(k_{\perp},k_{\|}\to {\bf G})=
\frac{k_{\perp}}{\pi^2kG^2}\int d^3k'\,\Pi_{l,t}({\bf G})
\eeq
is the dimensionless factor.

The transition probabilities (\ref{7n.4}) and (\ref{8m.3})
are quite general and fully defined but rather
complicated even for numerical analysis. However, a
drastic simplification can be achieved by analytic calculations even for the
general case.

In our present publication we consider
only a specific case that is when initial neutron energy is below the
optical  potential. Moreover, for the sake of simplicity  in the main
part of the paper we shall use only the results of the theoretical
calculations referring to an Appendix, where the calculations are
presented in details. Below we explain just the main steps of the
solution.

First, in A.1 we calculate the integral (\ref{5.5}). It turns out to
be performed "almost exactly" by transforming (\ref{5.5}) into a
contour integral in the complex $q_{\perp}$ plane with the use of the
four-pole structure of the
correlation function (\ref{3n.19.2}).

The next step, calculation of the sum in (\ref{5.3}), is performed in
A.2. Comparative importance of the 16 terms in the sum is very
sensitive to the position of the initial and the final neutron
energies. The result was obtained for the initial neutron
energy below the barrier and for any final energy of the scattered
neutron. After this
stage, the inelastic cross section is defined as a function of final neutron
momentum, ${\bf k}'$.

The last stage, integration over angular distribution of the
scattered neutrons, is done in A.3 and A.4. Here the momentum
transfer between the neutron and crystal lattice is very specific and
has required a quite different approach (see A.4).

We write down the final results for the dimensionless quantities
(\ref{7n.5}) and (\ref{8m.4}).

For transitions inside the first Brillouin zone (${\bf G}=0$) the
factor (\ref{7n.5}) was found as
\beq{8.45m}
\begin{array}{l}
W_l(k_{\perp},k_{\|}\to\ve')=
\Frac{k_{\perp} f(k'_{\perp})x_0}{3\pi k}+{}
\\[\bigskipamount]
\phantom{W_l(k_{\perp},k_{\|}\to\ve')}+
\Frac{2k_{\perp}\Gamma^2k'^2}
{\pi k\kp \left(k'^2-\epsilon\,\omega^2/c^2_l\right)^{3/2}},
\end{array}
\eeq
\beq{8.46m}
W_t(k_{\perp},k_{\|}\to\ve')=
\frac{2k_{\perp}f(k'_{\perp})x_0}{3\pi k},
\eeq
where
\beq{8.46m.2}
f(k'_{\perp})=\left\{
\begin{array}{ll}
4k'_{\perp}/k_0^2, & k'_{\perp}<k_0,\\
4/(k'_{\perp}+\sqrt{k'^2_{\perp}-k^2_0}\,), &
k'_{\perp}>k_0,
\end{array}\right.
\eeq
with $k'_{\perp}=\sqrt{k'^2-k^2_{\|}}$,
\beq{8.46m.3}
\kp=\sqrt{k_0^2-k_{\perp}^2},
\eeq
and
\beq{8.47m}
x_0(\omega)=\sqrt{\frac{\sqrt{(\epsilon\,\omega^2/c^2)^2+\Gamma^4}+
\epsilon\,\omega^2/c^2}{2}}.
\eeq

The factors (\ref{8m.4}) for the case ${\bf G}\ne 0$ were reduced to
\beq{8m.21m}
W_l(k_{\perp},k_{\|}\to{\bf G})=
\frac{4k_{\perp}x_0}{3k\kp},
\eeq
\beq{8m.21m.2}
W_t(k_{\perp},k_{\|}\to{\bf G})=
\frac{8k_{\perp}x_0}{3k\kp}.
\eeq

The transition probabilities (\ref{8.45m}), (\ref{8.46m}) and
(\ref{8m.21m}), (\ref{8m.21m.2}) can be further simplified by
averaging over the neutron incident angle. There are two factors with
angular
dependence in (\ref{8.45m}), (\ref{8.46m}) and (\ref{8m.21m}), namely
$k_{\perp}/(k\,\kp)$ and $k_{\perp}f(k'_{\perp})/k$. Following
(\ref{7n.13.2}) we introduce
\beq{8.48}
\av{\frac{k_{\perp}}{k\kp}}=
\frac{1}{4k}A(\frac{\ve}{U}),
\eeq
where
\beq{8.49}
A(x)=\frac{1}{x}\left(
\arcsin\sqrt{x}-\sqrt{x(1-x)}\,\right),
\eeq
and
\beq{8.50}
\av{\frac{k_{\perp}f(k'_{\perp})}{k}}=
\frac{1}{k}\,B(\frac{\ve}{U},\frac{\ve'}{U}),
\eeq
where the function $B(x,x')$ has a different form in three regions of
the variables: $x'<1$, $1<x'<1+x$, $1+x<x'$, and can be presented as
\beq{8.51}
B(x,x')=C(x,\frac{x'}{x})-C(x,\frac{x'-1}{x}),
\eeq
with
\beq{8.52}
C(x,y)=x\int\limits_{{\rm min}(0,y)}^{{\rm min}(1,y)}
\sqrt{(1-z)(y-z)}\,\,dz.
\eeq
Note the two simple limits
\beq{8.53}
\begin{array}{lll}
B(x,x')\simeq x/2, &
\mbox{for} \quad & |x'-x|\ll 1,
\\[\medskipamount]
B(x,x')\simeq (1/3)\sqrt{x/x'},\quad &
\mbox{for} & x'\gg 1.
\end{array}
\eeq

The final expressions for $W_{l,t}$, defined in (\ref{7n.5}) and
(\ref{8m.4}), averaged over the angle of incidence, are of the form
\beq{8.54}
\begin{array}{l}
W_l(\ve\to\ve')=
\Frac{x_0}{3\pi k}\,B(\Frac{\ve}{U},\Frac{\ve'}{U})+{}
\\[\bigskipamount]
\phantom{W_l(\ve\to\ve')=\Frac{\pi}{3k}}+
\Frac{\Gamma^2k'^2}{2\pi k
\left(k'^2-\epsilon\,\omega^2/c^2_l\right)^{3/2}}\,A(\Frac{\ve}{U}),
\end{array}
\eeq
\beq{8.55}
W_t(\ve\to\ve')=
\frac{2 x_0}{3\pi k}B(\frac{\ve}{U},\frac{\ve'}{U}),
\eeq
\beq{8m.22}
W_l(\ve\to{\bf G})=
\frac{x_0}{3k}A(\frac{\ve}{U}),\quad
W_t(\ve\to{\bf G})=
\frac{2 x_0}{3k}A(\frac{\ve}{U}).
\eeq
The second terms both in (\ref{8.45m}) and (\ref{8.54}) are valid
only for large energy transfer (see (\ref{8.30}), (\ref{8.30.2}) and
text before).

\section{Physical application. Specific ultra\-cold neutron interaction
with solids and liquids
\label{s7}}

In the previous section, specific cases of general equations were
analyzed with the use of
mathematical criteria. In this section, we consider the problems from
a physical
point of view, making classification by the relaxation processes,
phonon-like or
diffusion-like types, and the range of the energy transfer, with a
special attention to the
region of small heating and cooling, as well as to upscattering region.
The substances under investigation are of crystalline, amorphous
and liquid type, commonly used in experiments.

There are two longitudinal modes, phonon-like and thermo-diffusion,
both for solids
and liquids, and only one transverse mode. The latter is of the pure
phonon type
for solids, and of the combined type for liquids, that transforms
from the diffusion
type at small energy transfer to that of phonon type at large energy transfer
(see Section~\ref{Correlation function}).

\subsection{Phonon-like mode, ${\bf G}=0$
\label{s7.1}}

Phonon-like modes involve low damping, i.e. $\Gamma_p\to 0$.
Therefore, one can neglect the last term in (\ref{8.54}) and use for
(\ref{8.47m}) $x_0(\omega)=|\omega|/c$. For solids, when ${\bf G}=0$,
the sum over longitudinal and transverse modes gives from
(\ref{8.54}) and (\ref{8.55})
\beq{10.1}
\frac{dw_p(\ve\to\ve')}{d\ve'}=
\frac{\beta_c|\omega|n_0(\omega)}
{\pi Mc^3}\,B(\frac{\ve}{U},\frac{\ve'}{U}),
\eeq
where $c$ is defined by
\beq{10.19}
\frac{3}{c^3}=\frac{1}{c^3_l}+\frac{2}{c^3_t}.
\eeq

For the low energy part of the spectrum one can use (\ref{3n.4}) and find
\beq{10.5}
\frac{dw_p(\ve\to\ve')}{d\ve'}=
\frac{\beta_ck_Tv_T}{2\pi Mc^3}\,
B(\frac{\ve}{U},\frac{\ve'}{U}),\quad
|\ve'-\ve|\ll T.
\eeq
This part of the spectrum is completely determined by the function
$B(x,x')$. Here and below we use the quantities
\beq{10.6}
k_T=\frac{\sqrt{2mT}}{\hbar},\quad
v_T=\sqrt{\frac{2T}{m}}.
\eeq

The contribution of inelastic processes with ${\bf G}=0$ to the total
probability of the UCN escape from the trap can be estimated by
integration of (\ref{10.1}) from the barrier $U$ up to the Debye
energy
\beq{8.22m}
\ve_D=\hbar c(6\pi^2n)^{1/3}.
\eeq
Since $U\ll \ve_D$, it gives
\beq{10.6.2}
w_p(\ve)=c_1(\frac{\ve_D}{T})\,
\frac{\beta_ck_T}{12\pi}\,
\frac{m}{M}\,\frac{vv^2_T}{c^3},
\eeq
where $v=\sqrt{2\ve/m}$ is the initial neutron velocity, and
\beq{10.6.3}
c_1(y)=\int\limits_0^y\frac{x^{1/2}dx}{e^x-1}
\mathrel{\mathop{\,\,\longrightarrow\,\,}\limits_{y\to\infty}}
2.315.
\eeq

For liquids, the contribution of the longitudinal pho\-non-like mode is
of the form (\ref{10.1}), (\ref{10.5}) and (\ref{10.6.2}), where
$1/c^3$ should be replaced by $1/3c_l^3$. The transverse mode for
liquids is considered below (see Sec.~\ref{s7.4}).

\subsection{Phonon upscattering, ${\bf G}\ne 0$
\label{s7.2}}

The main process for upscattering in crystals goes with momentum
transfer from the lattice. Assuming $\Gamma_p\to 0$ and summing over
longitudinal and transverse phonons, we obtain for the total
upscattering probability from (\ref{8m.3}) and (\ref{8m.22})
\beq{10.18}
w^{up}_p(\ve)=\frac{\beta_c\hbar^3}{4m^2Mc^3}\,
A(\frac{\ve}{U})
\sum_{{\bf G}_{\|},G_{\perp}>0}
G^4n_0(\frac{\hbar G^2}{2m}),
\eeq
where $c$ is given by (\ref{10.19}). Restriction in the sum with
$G_{\perp}>0$ is made since the summation over  backward and forward
scattering was performed earlier.

To estimate the probability we replace summation over $G_{\perp}>0$
and ${\bf G}_{\|}$ by integration
\beq{10.20}
\sum_{{\bf G}_{\|},G_{\perp}>0}\quad
\longrightarrow\quad
\frac{1}{2}\int\frac{d^3G}{(2\pi/a_0)^3},
\eeq
where $a_0=\sqrt[3]{M/\rho}$ is the lattice constant. Integral over
$\ve'=\hbar^2G^2\!/2m$ up to Debye energy gives
\beq{10.21}
w^{up}_p(\ve)=c_2(\frac{\ve_D}{T})\,
\frac{\pi\beta_ck_T}{4}
\left(\frac{a_0k_T}{2\pi}\right)^3
\frac{m}{M}\left(\frac{v_T}{c}\right)^{\!3}\!\!A(\frac{\ve}{U}),
\eeq
where
\beq{10.12.4}
c_2(y)=\int\limits_0^y\frac{x^{5/2}dx}{e^x-1}
\mathrel{\mathop{\,\,\longrightarrow\,\,}\limits_{y\to\infty}}
3.745.
\eeq

The high energy upscattering due to phonons at ${\bf G}\ne 0$, was
considered earlier \cite{Ign90,Gol91} by other approaches. The
probability (\ref{10.21}) coincides with the result, presented, for
example, in review paper \cite{Ign00}.

\subsection{Longitudinal thermo-diffusion mode, ${\bf G}=0$
\label{s7.3}}

The differential probability of inelastic scattering via this mode,
which follows from (\ref{7n.4}) and (\ref{8.54}) with $\epsilon=0$,
can be presented in the form
\beq{10.14}
\begin{array}{l}
\Frac{dw_d(\ve\to\ve')}{d\ve'}=
\Frac{\alpha\beta_c\Gamma_dn_0(\omega)}
{3\sqrt{2}\,\pi Mc_l^2}
\left(B(\Frac{\ve}{U},\Frac{\ve'}{U})+
\phantom{\lefteqn{Frac{3\hbar\Gamma_d}{2\sqrt{m\ve'}}}}\right.
\\[\bigskipamount]
\left.\phantom{
\Frac{dwd_d}{d}}+
\Frac{3\hbar\Gamma_d}{2\sqrt{m\ve'}}\,
A(\Frac{\ve}{U})\,\theta(\ve'\!-\ve)
\left(1-e^{-\ve'\!/\!10\ve}\right)\right),
\end{array}
\eeq
where the last term in (\ref{8.54}), valid only for large energy
transfer ($\ve'\gg\ve$), is written with factors $\theta(\ve'-\ve)$
and $(1-e^{-\ve'/10\ve})$, which provide attenuation at small $\ve'$
($\theta(x)=0$ for $x<0$ and $\theta(x)=1$ for $x>0$). The damping
parameter, $\Gamma^2_d$, given in (\ref{3n.14.2}), can be presented as
\beq{10.15}
\frac{\Gamma_d}{k}=
\left(\frac{|\ve'-\ve|}{d_D\ve}\right)^{1/2},
\eeq
where
\beq{10.15.2}
d_D=\frac{2mD_S}{\hbar}
\eeq
is the dimensionless thermo-diffusion coefficient.

For small energy transfer, $\ve'\sim\ve$, the dominating first term
in (\ref{10.14}) takes the form
\beq{10.16}
\frac{dw_d(\ve\to\ve')}{d\ve'}\simeq
\frac{\alpha\beta_ck_T}{6\pi Mc_l^2}\,
\frac{vv_T}{v_0^2\sqrt{2d_D}}\,
\sqrt{\frac{\ve}{|\ve'-\ve|}},
\eeq
where $v_0=\sqrt{2U/m}$ is the barrier velocity. It is clear, that
the divergence at $\ve'\to\ve$ is integrable. This result for
longitudinal thermo-diffusion and for fixed components $k_{\perp}$
and $k_{\|}$ of the initial neutron momentum was obtained in
\cite{Bar00}.

For large energy transfer, the main contribution comes from the
second term in (\ref{10.14}). It should be remembered that diffusion
modes appear in hydrodynamic approximation. So, the energy transfer
via this mode has sense only inside the hydrodynamic region
(\ref{3n.12}), where $|\omega|\ll c^2_l/D_S$. It gives the boundary
for $\ve'$, where (\ref{10.14}) is reasonable,
\beq{10.16.2}
\ve'\ll\ve_d\equiv\frac{2mc_l^2}{d_D}.
\eeq
If $\ve_D<\ve_d$, the integration can be performed up to the Debye
energy. It gives
\beq{10.17}
w_d(\ve)=c_1(\frac{\ve_D}{T})\,
\frac{\alpha\beta_ck_T}{4\pi d_D}\,
\frac{m}{M}\!\left(\frac{v_T}{c_l}\right)^{\!\!2}\!\!A(\frac{\ve}{U}).
\eeq

\subsection{Transverse mode for liquids, ${\bf G}=0$
\label{s7.4}}

Taking $\alpha=\epsilon=1$ (see text after (\ref{3n.19.2})), we get
from (\ref{7n.4}) and (\ref{8.55}) the differential probability of
inelastic scattering
\beq{10.1t}
\frac{dw_t(\ve\to\ve')}{d\ve'}=
\frac{2\beta_cx_0(\omega)n_0(\omega)}
{3\pi Mc_t^2}\,
B(\frac{\ve}{U},\frac{\ve'}{U}),
\eeq
where $x_0(\omega)$ is given by (\ref{8.47m}) with $c\to c_t$, while
the damping factor $\Gamma^2\to \Gamma_t^2$ is of the form
(\ref{3n.18}).

Hydrodynamic region (\ref{3n.18.2}), $|\omega|\ll c^2_t/\nu$,
corresponds to the final neutron energy
\beq{10.8.2}
\ve'\ll\ve_{\nu}\equiv\frac{2mc_t^2}{d_{\nu}},
\eeq
where
\beq{10.8.3}
d_{\nu}=\frac{2m\nu}{\hbar}
\eeq
is the dimensionless viscosity. On the other hand, (\ref{10.8.2}) means
$\Gamma_t^2=|\omega|/\nu\gg\omega^2/c_t^2$, which simplifies $x_0(\omega)$,
\beq{10.8.4}
x_0(\omega)=\frac{\Gamma_t}{\sqrt{2}}.
\eeq

In the region $\ve'\sim\ve$, (\ref{10.1t}) takes the form
\beq{10.9}
\frac{dw_t(\ve\to\ve')}{d\ve'}\simeq
\frac{\beta_ck_T}{3\pi Mc_t^2}\,
\frac{vv_T}{v_0^2\sqrt{2d_{\nu}}}\,
\sqrt{\frac{\ve}{|\ve'-\ve|}},
\eeq
similar to (\ref{10.16}) for longitudinal thermo-diffusion.

Beyond the hydrodynamic region, i.e. for large energy transfer, we have
\beq{10.9.2}
\Gamma_t^2\ll\frac{\omega^2}{c_t^2}
\quad\Longrightarrow\quad
x_0(\omega)=\frac{|\omega|}{c_t},
\eeq
and (\ref{10.1t}) transforms into (\ref{10.1}) for transverse
phonon-like mode, where $1/c^3$ should be replaced by
$2/3c_t^3$.

\subsection{Relative contribution of phonons and thermo-diffusion to
large energy transfer
\label{s7.5}}

Phonon contribution to large energy transfer essentially depends on
momentum exchange with the lattice. The probability for upscattering
with (${\bf G}\ne 0$, (\ref{10.21})) and without (${\bf G}=0$,
(\ref{10.6.2})) such exchange differ by the factor $v/c$ ($\sim
10^{-3}$ for UCN). The physical reason is simple. Inelastic
scattering can be visualized as an absorption by the neutron of a
quantum from the media with the energy $\hbar\omega$ and momentum
$\hbar q$. The requirement for energy and momentum conservation does
not allow this process with "free" quanta. But for UCN the normal
component of momentum is pure imaginary, $i\kp$, and instead of a
$\delta$-function, as in (\ref{3.9}) for ${\bf q}_{\|}$, one has from
(\ref{3.11})--(\ref{3.12}) for $q_{\perp}$
\beq{6m.3.2}
\Sigma^{1\sigma'}_{\perp}\sim
\frac{e^{i(q_{\perp}+G_{\perp}-\sigma'\bar k')a/2}}
{\kp+i(q_{\perp}+G_{\perp}-\sigma'\bar k')}.
\eeq

Integrals over $q_{\perp}$ with (\ref{6m.3.2}) are very sensitive to
the minimum value of imaginary factor,
$q_{\perp}+G_{\perp}-\sigma'\bar k'$, which may be reached in the
allowed $q_{\perp}$ range. For sound-like excitations, a rough
estimate gives
\beq{6m.3.3}
q=\frac{\omega}{c}\sim\frac{v'}{c}\,k'\ll k',
\eeq
where $v'$ and $k'$ are the velocity and wave number of scattered
neutron. It is evident, that the large factor $\sigma'k'$ in the
denominator of (\ref{6m.3.2}) can be compensated only by $G_{\perp}$.
Then (\ref{6m.3.2}) can result in the large factor $1/\kp$, which is
seen in (\ref{8m.21m}), (\ref{8m.21m.2}). With $G_{\perp}=0$ one may
expect, instead of $1/\kp$, the factor $1/k'_{\perp}$, the result at
least by the factor $v/c$ smaller.

Note, that the factor $1/\kp$ is of the scale of the UCN intrusion
length into the wall of the trap. Therefore, it is natural that the
inelastic scattering probability is proportional to $1/\kp$. Thus,
the factor $1/k'_{\perp}$ instead of $1/\kp$ points to substantial
suppression of the large energy transfer by the phonon mode with
${\bf G}=0$.

The thermo-diffusion mode can be also effective for the large energy
transfer, but the physics is quite different. Parameters $\omega$ and
$q$ of "diffusion quanta" are not strongly coupled, as in the phonon
case, (\ref{6m.3.3}). Even with $\omega$ fixed by the energy
transfer, $q_{\perp}$ is still allowed to be varied up to $\sim k'$.
Thus, (\ref{6m.3.2}) can reach $1/\kp$ even for ${\bf G}=0$. The
result can be seen from the last term in (\ref{8.45m}). It vanishes
in the limit $\Gamma\to 0$ (for phonons), but is proportional to
$1/\kp$ for thermo-diffusion and results in the probability for
upscattering (\ref{10.17}) which may be comparable to that for
phonons with ${\bf G}\ne 0$, (\ref{10.21}).

\section{Results for two specific target samples
\label{s8}}

In previous section we have obtained analytic results for each type
of relaxation modes. To get the physical result for a specific
experiment, one should combine several analytic formulae with the
relaxation modes and parameters appro\-pri\-ate\-ly chosen. To demonstrate
this procedure and examine general trends of final results and their
sensitivity to physical parameters, it is worthwhile to consider some
specific examples.

We have chosen stainless steel (SS) and Fomblin fluid (FF), two
materials which are often used in real experiments and, on the other
hand, have quite different relaxation properties. Both materials are
not of a simple structure, and we shall use for them simplified
descriptions, which are specified below.

\begin{table*}
\caption{Parameters for stainless steel (SS) and Fomblin fluid (FF).}
\label{tab:1}       

\begin{tabular}{|c|c|c|c|c|c|c|c|c|c|c|}
\hline\lefteqn{\phantom{\frac{\strut 1}{\strut 2}}}
& $T$, K & $K$, Pa & $\beta_P$, K$^{-1}$ & $C_P$,
$\frac{\mbox{J}}{\mbox{g}\cdot\mbox{\strut K}}$ &
$\kp$, $\frac{\mbox{W}}{\mbox{m}\cdot\mbox{\strut K}}$ &
$\nu$, $\frac{\mbox{cm}^2}{\mbox{\strut sec}}$ & $\alpha$ &
$D_S$, $\frac{\mbox{cm}^2}{\mbox{\strut sec}}$ & $d_D$ & $d_{\nu}$ \\
\hline\lefteqn{\phantom{\frac{1}{2}}}
SS & 293 & 170$\cdot$10$^9$ & 3.6$\cdot$10$^{-5}$ & 0.45 & 80 & 0.01 & 0.018 &
2.28$\cdot$10$^{-1}$ & 717.5 & 31.5 \\
\hline\lefteqn{\phantom{\frac{1}{2}}}
FF & 293 & 3$\cdot$10$^9$ & 6$\cdot$10$^{-4}$ & 1.0 & 0.2 &
1.40 & 0.167 & 1.05$\cdot$10$^{-3}$ & 3.30 & 4405.7 \\
\cline{2-11}\lefteqn{\phantom{\frac{1}{2}}}
& 373 & 2$\cdot$10$^9$ & 7$\cdot$10$^{-4}$ & 1.1 & 0.2 &
0.07 & 0.175 & 0.96$\cdot$10$^{-3}$ & 3.02 & 220.3 \\
\hline
\end{tabular}

\end{table*}

\subsection{Parameters for stainless steel}

The nuclei $^{56}$Fe are the dominant ones in stainless steel. Thus,
we take $A=56$, where
$A=M/m$ is the mass number of the scatterer. Then,
$\rho=7.8$~g/cm$^3$ and, therefore, $n=8.4\cdot 10^{22}$~cm$^{-3}$
and $a_0=2.3\cdot 10^{-8}$~cm. For the coherent scattering length we
take $b_c=0.87\cdot 10^{-12}$~cm, which gives for the barrier energy
$U=192$~neV. Sound
velocities at $T=293$~K are equal to: $c_l=5850$~m/sec and
$c_t=3230$~m/sec.

For the case of stainless steel one should consider contributions
from longitudinal and transverse phonons and thermo-diffusion, the
main part of the diffusion-like processes. For the latter we need
$\alpha=C_P/C_V-1$ and the thermo-diffusion coefficient $D_S$,
\beq{10c.1}
\alpha=\frac{TK\beta^2_P}{\rho C_P},\quad
D_S=\frac{\kp}{\rho C_P},
\eeq
where $K=-V(\partial P/\partial V)$ is the volume elasticity, and
$\beta_P=(\partial V/\partial T)_P/V$ is the thermal coefficient of
volume expansion. The corresponding values at normal temperature
$T=293$~K are listed in Table~1 as well as the calculated quantities
$\alpha$, $D_S$, and dimensionless parameters $d_D$ and $d_{\nu}$.

\subsection{Parameters for Fomblin fluid}

Fomblin is a hydrogen-free fluorinated oil (it is known also as
perfluropolyether). It is quite viscous at
room temperature and used to cover the walls of UCN storage traps
because of the small neutron absorption and upscat\-te\-ring rate. There
are different modifications  of Fomblin oil and not all of them are
described in the literature in detail. Thus we use typical values for
Fomblin parameters. In addition, we ignore complex structure of
Fomblin molecules and are treating it as a sim
ple liquid.

The stoichiometry is roughly C$_3$F$_6$O. Thus
we assume that the $^{16}$F nucleus dominates in the fluid and
$A=16$ is the mass number of scatterer. We take
$\rho=1.9$~g/cm$^3$, then $n=7.15\cdot 10^{22}$~cm$^{-3}$. For
the coherent scattering length we use $b_c=0.565\cdot 10^{-12}$~cm,
which gives $U=106$~neV in accordance with experiment. For all
temperatures we take fixed sound velocities: $c_l=1900$~m/sec and
$c_t=1500$~m/sec.

Parameters $K$, $\beta_P$, $C_P$, $\kp$, $\nu$
as well as the calculated $\alpha$, $D_s$, and dimensionless quantities
$d_D$ and $d_{\nu}$ are presented in Table~1
for two temperatures: $T=293$~K and 373~K. Among these parameters
viscosity $\nu$ is the most sensitive to the temperature.

\subsection{Numerical results}

We have calculated the probability of inelastic scattering for UCN
with initial energy near half of the barrier height, that is for
$\ve=80$~neV for stainless steel and $\ve=40$~neV for Fomblin.
Contributions of phonons and diffusion modes are shown separately.
\begin{figure}
\begin{center}
\mbox{\includegraphics*[scale=0.6]{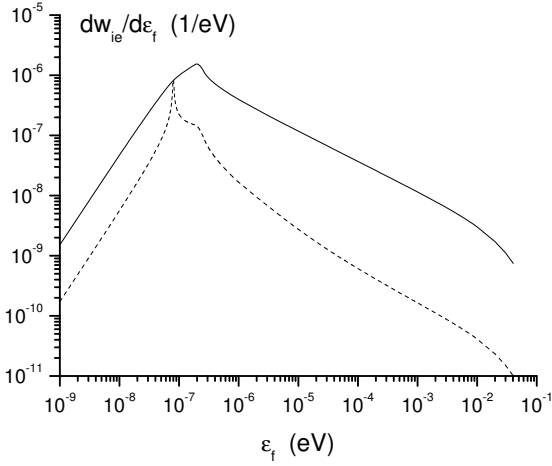}}
\caption{\label{F1} Inelastic scattering for UCN of 80 neV on stainless steel
at $T=293$~K: the
differential probability $dw_{ie}/d\ve_f$ per bounce as function of the final
neutron energy $\ve_f$. Solid line -- phonon contribution, dashed line --
thermo-diffusion contribution.}
\end{center}
\end{figure}
\begin{figure}
\begin{center}
\mbox{\includegraphics*[scale=0.6]{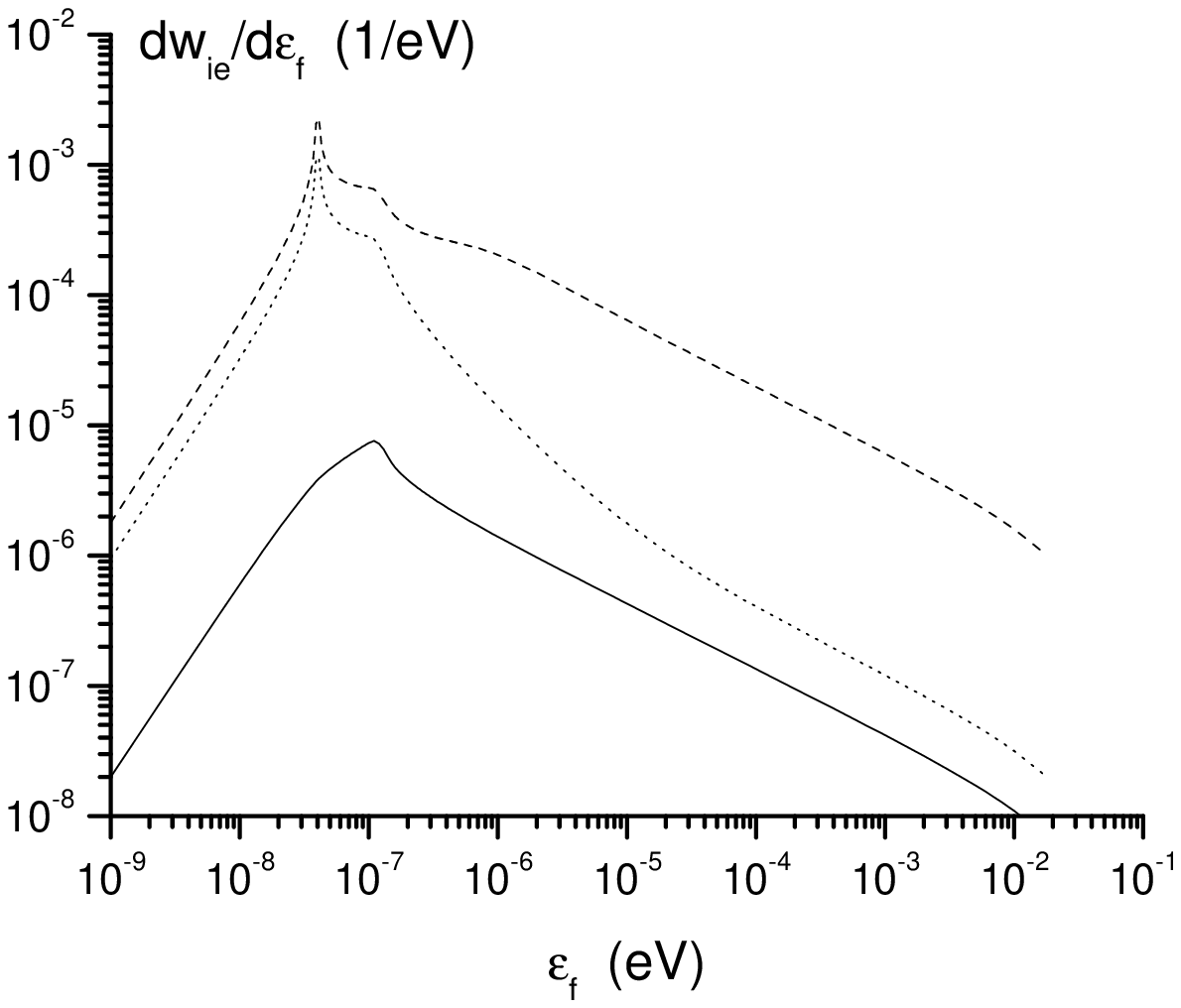}}
\caption{\label{F2} Inelastic scattering for UCN of 40 neV on Fomblin oil
at $T=293$~K: the
differential probability $dw_{ie}/d\ve_f$ per bounce as function of the final
neutron energy $\ve_f$. Solid line -- longitudinal sound contribution, dotted
line -- transverse sound contribution, dashed line --
thermo-diffusion contribution.}
\end{center}
\end{figure}
\begin{figure}
\begin{center}
\mbox{\includegraphics*[scale=0.6]{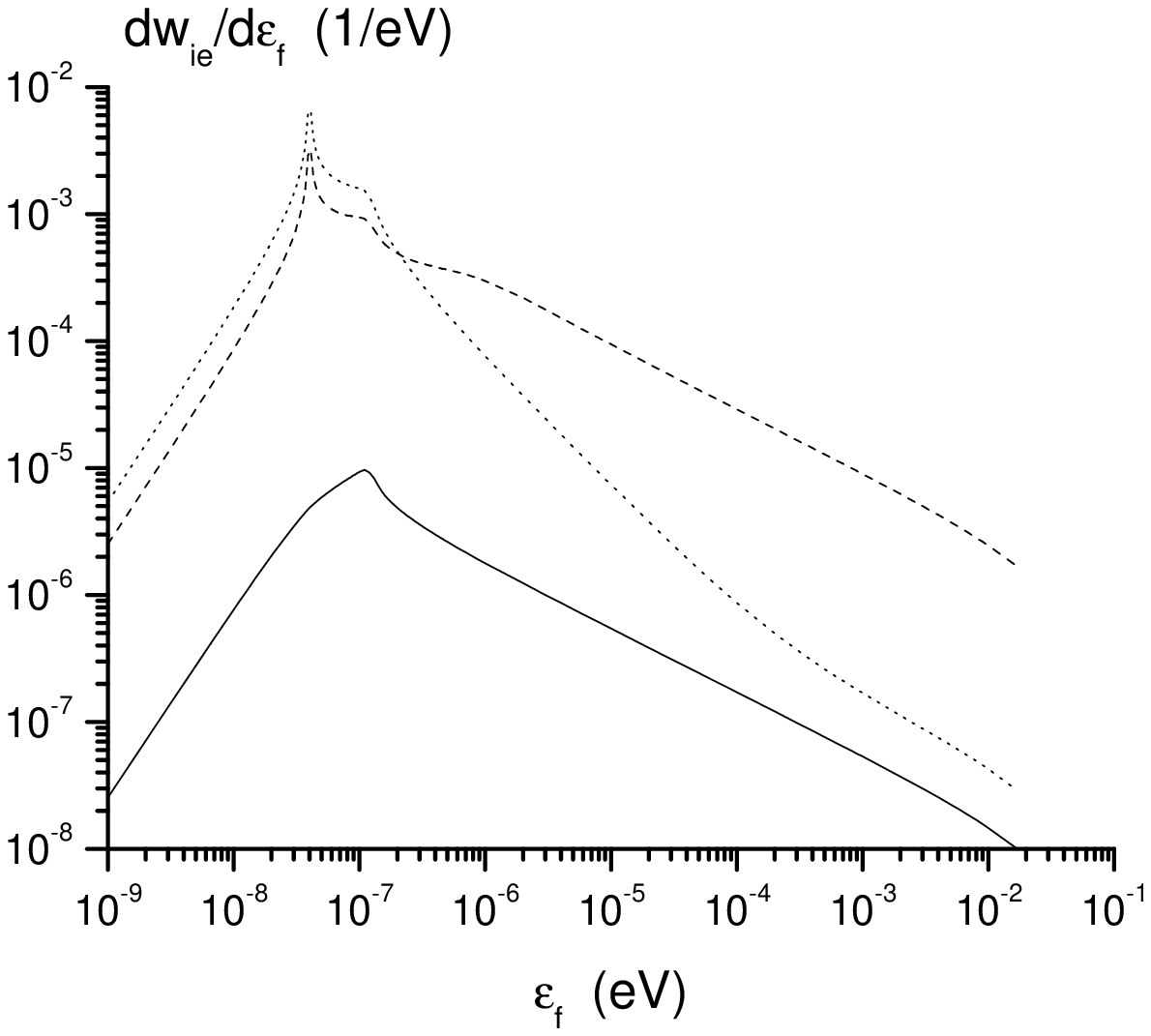}}
\caption{\label{F3} Inelastic scattering for UCN of 40 neV on Fomblin oil
at $T=373$~K: the
differential probability $dw_{ie}/d\ve_f$ per bounce as function of the final
neutron energy $\ve_f$. Solid line -- longitudinal sound contribution, dotted
line -- transverse sound contribution, dashed line --
thermo-diffusion contribution.}
\end{center}
\end{figure}
Results for the differential probability are presented on Figs.~\ref{F1}, \ref{F2}
and \ref{F3}.

\begin{figure}
\begin{center}
\mbox{\includegraphics*[scale=0.6]{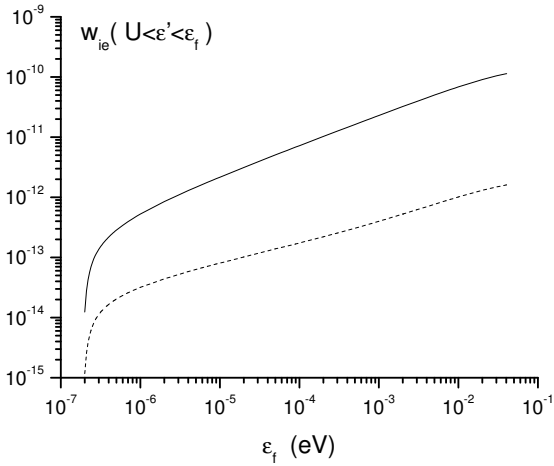}}
\caption{\label{F4} Inelastic scattering for UCN of 80 neV on stainless steel
at $T=293$~K: the
integral probability $w_{ie}$ per bounce of transition to the interval
$U<\ve<\ve_f$ as function of $\ve_f$. Solid line -- phonon
contribution, dashed line -- thermo-diffusion contribution.}
\end{center}
\end{figure}

\begin{figure}
\begin{center}
\mbox{\includegraphics*[scale=0.6]{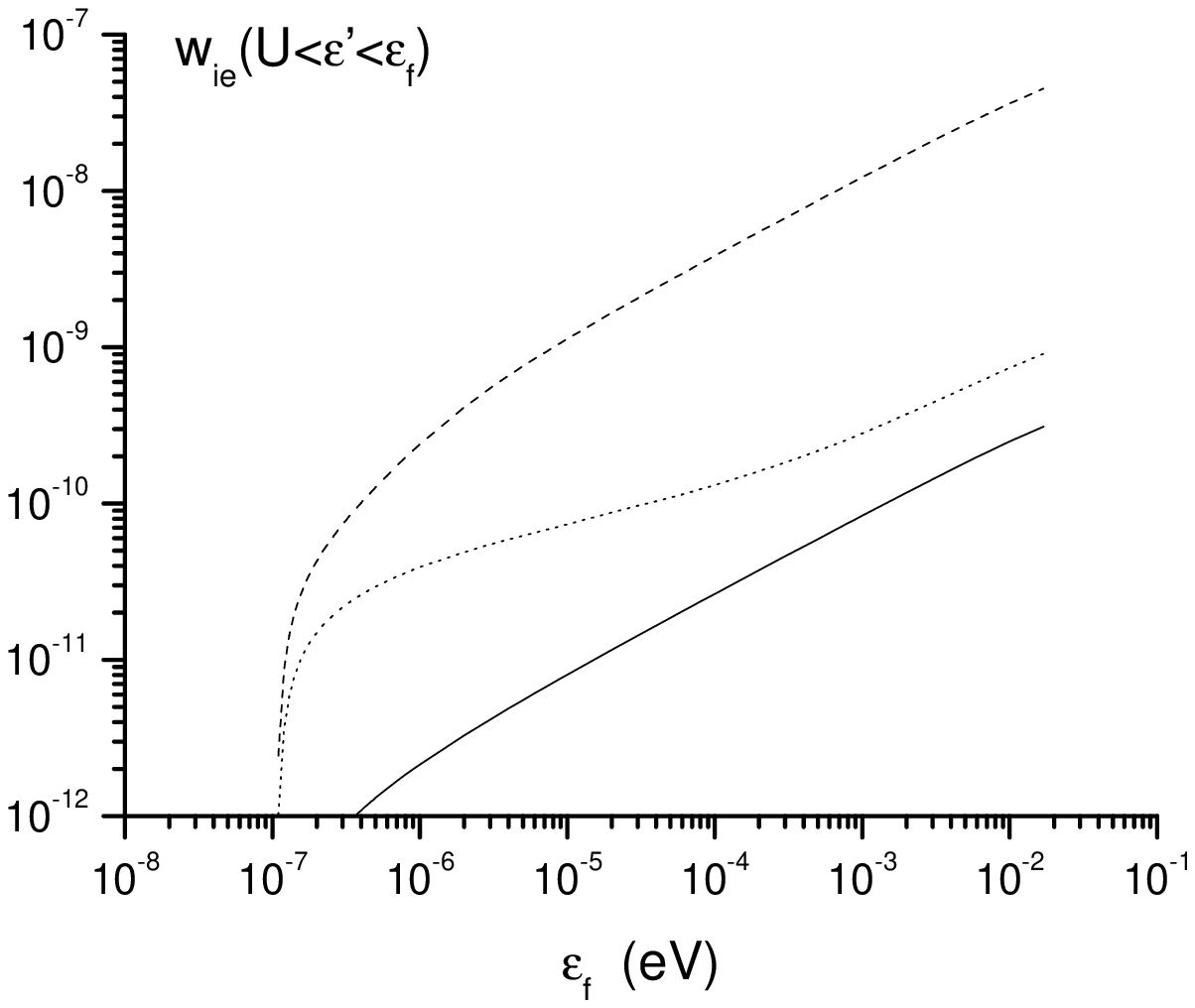}}
\caption{\label{F5} Inelastic scattering for UCN of 40 neV on Fomblin oil
at $T=293$~K:  the
integral probability $w_{ie}$ per bounce of transition to the interval
$U<\ve<\ve_f$ as function of $\ve_f$. Solid line -- longitudinal
sound contribution, dotted line -- transverse sound contribution,
dashed line -- thermo-diffusion contribution.}
\end{center}
\end{figure}

\begin{figure}
\begin{center}
\mbox{\includegraphics*[scale=0.6]{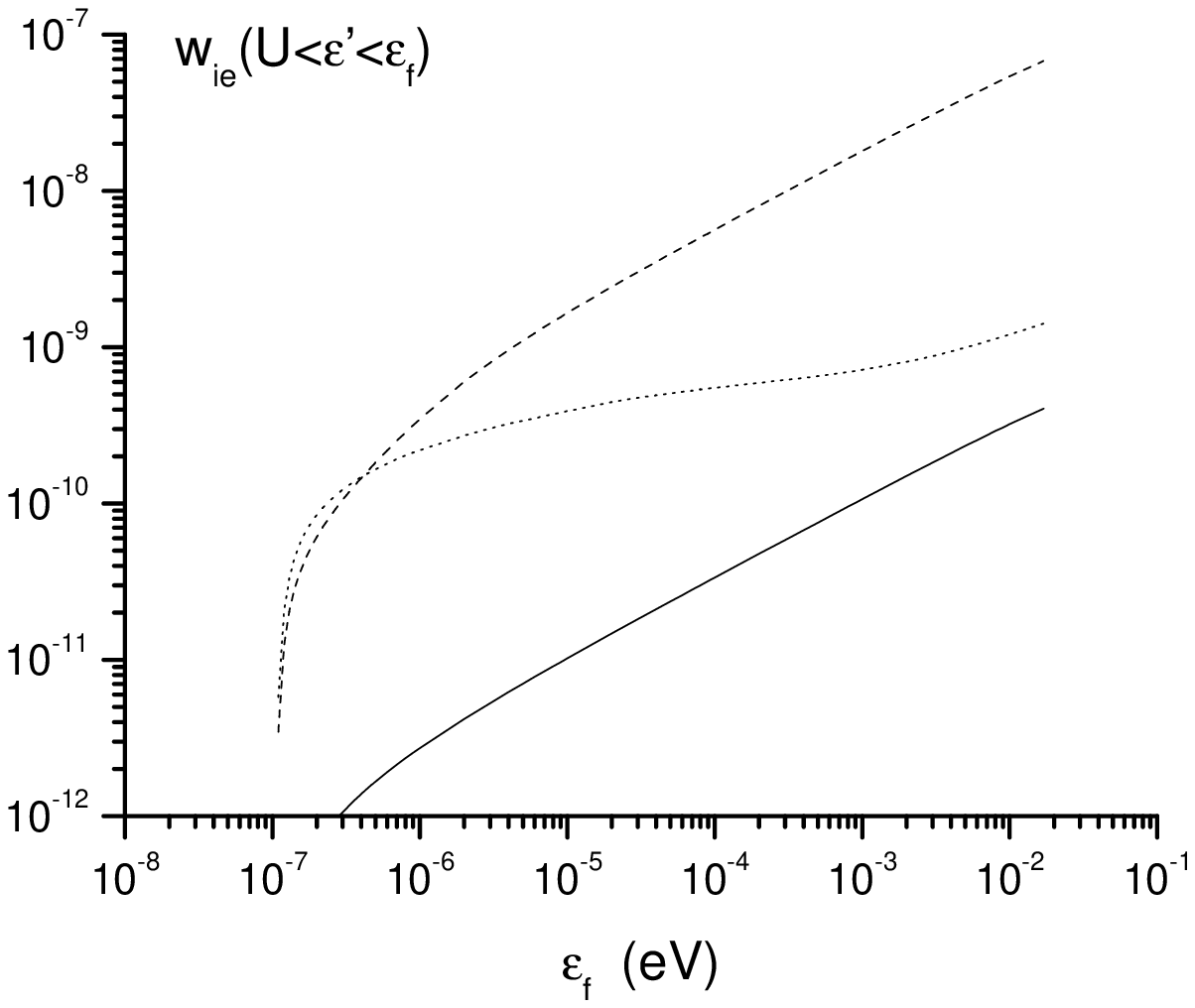}}
\caption{\label{F6} Inelastic scattering for UCN of 40 neV on Fomblin oil
at $T=373$~K: the
integral probability $w_{ie}$ per one bounce of transition to the interval
$U<\ve<\ve_f$ as function of $\ve_f$. Solid line -- longitudinal
sound contribution, dotted line -- transverse sound contribution,
dashed line -- thermo-diffusion contribution.}
\end{center}
\end{figure}

Integral probabilities
\beq{10c.2}
w(\ve_f)=\int\limits_U^{\ve_f}\frac{dw(\ve\to\ve')}{d\ve'}\,d\ve'
\eeq
of transition to the interval $U<\ve'<\ve_f$ per bounce are shown on
Figs.~\ref{F4}, \ref{F5} and \ref{F6}.
For Fomblin both differential and integral
inelastic scattering probabilities are presented for two different
temperatures: 293~K and 373~K.

Test calculations have been carried out starting from the exact
expressions (\ref{7n.4}) and (\ref{7n.5}). Comparison with results
from simplified analytic formulae has shown quite small deviations in
the whole energy range. So, we present numerical results based on
analytic formulae.

For stainless steel, the phonon contribution at ${\bf G}=0$ was
calculated from (\ref{10.1}) and thermo-diffusion contribution --
from (\ref{10.14}). Results at $T=293$~K are shown by solid and
dashed lines, respectively, on Figs.~\ref{F1} and \ref{F4}. The final neutron
energy for stainless steel is restricted from above by the Debye
energy $\ve_D=0.040$~eV (\ref{8.22}) with sound velocity $c$ from
(\ref{10.19}). The thermo-diffusion contribution should be restricted
from above by the limit energy $\ve_d=0.001$~eV (see (\ref{10.16.2})).

Integral probabilities on Fig.~\ref{F4}, calculated for ${\bf G}=0$, should
be compared with the total upscattering probability at ${\bf G}\ne
0$, which is given by (\ref{10.21}). At the temperature $T=293$~K, it
is equal to
\beq{10c.3}
w^{up}_p=0.7\cdot 10^{-6}.
\eeq

For Fomblin, the longitudinal sound (phonon) contribution is given by
(\ref{10.1}) with $1/c^3\to 1/3c_l^3$ (see the end of
Sec.~\ref{s7.1}), the transverse sound contribution -- by
(\ref{10.1t}) and thermo-diffusion contribution -- by (\ref{10.14}).
They were calculated both at $T=293$~K and $T=373$~K and are shown by
solid, dotted and dashed  lines, respectively, on Figs.~\ref{F2}, \ref{F3}, \ref{F5} and
\ref{F6}. The final neutron energy is restricted from above by the Debye energy
$\ve_D=0.017$~eV (\ref{8.22}) with sound velocity from (\ref{10.19}).
For Fomblin, the energy $\ve_d=0.023$~eV (\ref{10.16.2}), the limit
for thermo-diffusion contribution, is greater than the Debye energy
$\ve_D$.

\section{Discussion \label{Discussion}}

Differential spectra (Figs.~\ref{F1}, \ref{F2} and \ref{F3}) have well developed maxima
for both modes, phonon (with ${\bf G}=0$) and diffusion-like. For the
phonon mode, the peak shape is determined by the function
$B(\ve/U,\ve'/U)$. It has a maximum at $\ve'=U$ ($\ve<U$). In
diffusion-like modes there is a divergence (integrable) at
$\ve'\simeq\ve$ (see (\ref{10.16}) and (\ref{10.9})). These general
forms of the curves are the same for stainless steel and Fomblin, but
the relative contribution from phonon-like and diffusion-like modes
are quite different for the two substances. Contributions from
phonons and diffusion into small energy transfer can be estimated
from (\ref{10.5}) and (\ref{10.16}), (\ref{10.9}). Omitting numerical
factors, one has for diffusion-to-phonon contribution ratio
\beq{20.1}
\frac{w_d}{w_p}\sim
\frac{\alpha c}{v_0\sqrt{d}}.
\eeq

For thermo-diffusion, the high value of the ratio $c/v_0$ may be
compensated by the smallness of the quantity $\alpha=C_P/C_V-1$ and
additionally suppressed by the factor $1/\sqrt{d_D}$, when $d_D$ is
large. This is just the case for stainless steel, when the right part
of (\ref{20.1}) is near unity. As for Fomblin, the right-hand side of
(\ref{20.1}) is near 40. This qualitative estimate agrees with
numerical results presented on Figs.~\ref{F1}, \ref{F2} and \ref{F3}.
The thermo-diffusion
contribution is comparable with the phonon one for stainless steel
(Fig.~\ref{F1}) and dominates for Fomblin fluid (Figs.~\ref{F2} and \ref{F3}).

The transverse relaxation mode in liquids for small energy transfer
is also of diffusion type, which originates from sound waves with
high damping due to viscosity. The contribution of this mode, as
compared to the longitudinal phonon one, can be evaluated from
(\ref{20.1}) with $\alpha=1$. It far exceeds unity in agreement with
Figs.~\ref{F2} and \ref{F3}.

It should be emphasized that the two diffusive modes, which, due to
the large ratio (\ref{20.1}), give the main contribution to the small
energy transfer for Fomblin, are governed by different physical
parameters. The transverse mode has no small parameter $\alpha$, but
its quantity $d\to d_{\nu}$ (\ref{10.8.3}) differs from that for
thermo-diffusion, $d_D$ (\ref{10.15.2}), by magnitude and temperature
dependence (see Table 1).

{\it Remark.} The transverse mode was considered here with a simple
liquid model (\ref{3n.15})--(\ref{3n.18}), where transition to
"ideal" liquid with very low viscosity is not correct. In this paper
we have performed calculations for $d\ge 1$ (see Sec.~\ref{sA.3.1}),
so the definition (\ref{10.8.3}) gives the lower bound for the
viscosity $\nu$. (In \cite{Bar00} we have argued, that for small
$d<1$ the contribution of diffusion mode to inelastic scattering
vanishes.)

Let us now discuss quantitative results for the probability per
bounce (\ref{10c.2}) for the neutron to change its energy from $U$ up
to $(10-10^3)\cdot U$, that is to go from ultra cold to very cold
energy region ("small heating"). For stainless steel,
the probability for neutron transition to the region
from $U$ to $(10-10^3)\cdot U$ is of the scale of $\sim 10^{-13}-10^{-11}$
(Fig.~\ref{F4}), while for Fomblin it is of the scale of
$\sim 10^{-10} - 10^{-9}$ (Figs.~\ref{F5} and \ref{F6}). Evidently,
these contributions could not be seen in the measurements
\cite{Str00,Bon02,Lyc02,Ste02,Ser03}.
Nevertheless note, that our calculations
for Fomblin shows a definit increase of low-energy heating
with the temperature due to the contribution from diffusion-like
transverse mode. It is in a qualitative agreement with the data
\cite{Bon02,Ser03}. The measured effect of small heating on Fomblin seems
to be explained by surface effects (see \cite{Ste02,Ser03}).

As a general conclusion, phonons are ineffective for small energy
transfer both for stainless steel and Fomblin. Ther\-mo-diffusion is
even less effective for stainless steel, but gives very large effect
(one-two orders of magnitude larger than phonons) for Fomblin, where
the transverse diffusion mode gives comparable contribution and even
exceeds thermo-diffusion when the temperature goes up. (Note,
however, that it goes down more steeply when the transferred energy
becomes higher.)

In the region of high energy transfer, the ratio of thermo-diffusion
and phonon (at ${\bf G}=0$) contributions can be estimated from
(\ref{10.6.2}) and (\ref{10.17}),
\beq{20.2}
\frac{w_d}{w_p}\sim
\frac{\alpha c}{v_0d_D}.
\eeq
It is of the scale of 10$^{-1}$ for stainless steel and 10$^2$ for
Fomblin fluid. Thus, thermo-diffusion may dominate over phonons not
only for small energy transfer, but as well for a large one, provided
the dimensionless thermo-diffusion coefficient $d_D$ is low. This conclusion
may have sense only for liquids and amorphous materials, since for
crystals, and stainless steel in particular, the main process for the
transition up to the temperature region, upscattering, is that with
momentum exchange with the lattice, ${\bf G}\ne 0$ (see
Sec.~\ref{s7.5}).

Therefore, the processes responsible for high energy transfer are
quite different for crystals and amorphous materials.
Our result for coherent UCN
upscattering on pure crystalline material (\ref{10.21}) coincides
with that previously obtained. It is known that
its estimate (\ref{10c.3}) is one-two
orders of magnitude too small to be observed in a real storage
experiments. The reason is, evidently, a much higher absorption rate due to the
radiative capture and possible upscattering on hydrogen contamination
of the surface.

For Fomblin only the surface effects have been
discussed in details so far. We have not found any
published results with evaluation of escape probability from the bulk
properties of this material. In other words, there is no understanding
why the parameter $f$ (see \cite{Lam02}) of bulk losses lies in the range
10$^{-5}$ to 10$^{-6}$. Thus our result just for bulk effect given by
the integral (\ref{10c.2}) (calculated with $\ve_f=\ve_D$) is new.
Numerical results for Fomblin
presented in Figs.~\ref{F5} and \ref{F6} give for the loss probability
per bounce ($\simeq f$) the following values: $0.5\cdot 10^{-7}$ at 293~K and $0.7\cdot
10^{-7}$ at 373~K. Possible reasons for the
discrepancy of one-two orders of magnitude are mentioned in conclusion.

\section{Conclusion and outlook
\label{s9}}

A theory for coherent inelastic scattering of ultracold neutrons was
developed with a quite general form of the correlation function,
which allows to consider energy-momentum exchange of the
neutron with sound waves (phonons) as well as with diffusion-like
relaxations. The latter appears when a correlation function is
considered in the hydrodynamic region of momentum
and energy transfers (the low limit for ${\bf q}$ and $\omega$). Namely,
we discussed the longitudinal thermo-diffusion mode both for
solids and liquids as well as the diffusion-like  transverse mode
for liquids.

In the frame of this theory, we evaluated the
probability per bounce for an ultracold neutron to gain energy and to
go into cold or thermal region. In this paper the case ${\bf G}=0$
(i.e. no momentum transfer between neutron and the crystal lattice)
is consistently considered for the first time.

Numerical estimates are presented for the stainless steel storage
bottle and one with Fomblin-fluid coated walls. For Fomblin, the most
explored experimentally, theory predicts the overwhelming effect for
thermo-diffusion processes, two-three orders of magnitude larger than
from phonons, and a quite strong dependence on the temperature
originated from viscosity. This result is in a qualitative agreement
with experiments.

However, the numerical values
for the upscattering probability per bounce obtained for stainless steel
as well as for Fomblin are
still one-two order of magnitude too small to fully explain
experimental data. Thus, it is worthwhile to mention restrictions of
the presented results and possible future improvements.

Results from diffusion processes are quite sensitive to the
parameters of correlation function. It is very desirable to clarify
the structure and parameters of correlation function, especially for
liquids. In fact, it can be done by experiments in optics and other
means outside the neutron physics.

We considered our samples as a spatially uniform media. This may not
be correct for Fomblin, which exhibits some polymer features and may
have a non-uniform structure at the scales comparable with the
ultacold neutron reduced wave length $\sim 10$~nm.
To get the correlation function for
polymer and to include it in the neutron scattering theory is a
serious problem for the future. Nevertheless, the effect may be
evaluated from simple physical arguments. The probability for energy
transfer, as was mentioned above, is proportional to the intrusion
length or, what is the same, to the time
of neutron-wall collision. Spatial polymer
structure changes, in particular, the self-diffusion parameters
for the neutron and increases
the time, spent by the neutron inside the sample during the collision.
This time can be evaluated from transmission experiments and the
corresponding correction can be easily introduced into the formulae
of this paper.

On the other hand, gigantic polymer molecules can behave as
nano-particles. If it is the case, elastic scattering of the neutron by
the whole moving molecule in their center-of-mass system would
result in inelastic scattering of neutron in the laboratory
system. Similar mechanism of small cooling and heating due to
nano-particles at the surface of solid materials was proposed in
\cite{Nes02}.

No surface effect was included, which seems to be of no importance in
solids for UCN large wave length. This argument may not concern
specific surface excitations in liquids, say, capillary waves
\cite{Pok99b,Lam02}, which can be included into consideration as one
of the excitation modes. This particular case requires a different
correlation function and is the subject for a separate paper.

Several important effects were not included in this paper. Among them
are spin-flip effects and surface contamination with hydrogen
etc. Note, for example, that in \cite{Kor04} evidences are presented that the
total upscattering of UCN  is mostly related just to hydrogen
contamination of the walls of material traps. All these effects are of
importance for incoherent scattering, not included in this paper,
which deals only with coherent inelastic scattering of initially
ultracold neutrons.
\bigskip\bigskip

{\large\bf Acknowledgements}
\bigskip

We are kindly grateful to V.I.Morozov, E.Korobkina and R.Golub for valuable discussions.
The work was supported by RFBR grant 02-02-16808, grant NS-1885. 2003.2
and the program of bilateral
cooperation between Russia and Germany, Grant No. RUS 02/030.
\bigskip\bigskip

\appendix

\section{Calculation of inelastic cross section and inelastic
transition probability}

The aim of this Appendix is to simplify the general expression
(\ref{5.2})--(\ref{5.3})
for inelastic cross section. The first stage is calculation of the integral
(\ref{5.5}).

\subsection{Calculation of the integral $J_{l,t}(\lambda,\eta)$ (\ref{5.5})}

We start with a general integral of the form
\beq{A1}
I(\lambda,\eta)=
\frac{1}{\pi}
\int\limits_{-\infty}^{+\infty}
dq_{\perp}\,
\Delta(q_{\perp}-\lambda)
\Delta(q_{\perp}-\eta)
{\cal F}(q_{\perp}),
\eeq
where $\lambda$ and $\eta$ are complex
parameters, and ${\cal F}(q_{\perp})$ is a function with poles
above ($q_{n\uparrow}$) and below ($q_{n\downarrow}$) the real
axis, but not near it. The function $\Delta(x)$ is defined by
(\ref{3.12}), and therefore \beq{A2}
\begin{array}{l}
\Delta(q_{\perp}-\lambda) \Delta(q_{\perp}-\eta)=
\Frac{1}{(q_{\perp}-\lambda) (q_{\perp}-\eta)}\times{}
\\[\bigskipamount]
\phantom{}\times \left(2\cos\Frac{(\lambda-\eta)a}{2}-
e^{i\left(q_{\perp}-\efrac{\lambda+\eta}{2}\right)a}-
e^{-i\left(q_{\perp}-\efrac{\lambda+\eta}{2}\right)a}\right).
\end{array} \eeq

We transform (\ref{A1}) into the contour integral closing a path in the upper
or the lower half-plane. Then, neglecting the term proportional to
$e^{-{\rm Im}\,q_{n\uparrow}a}$ and $e^{-|{\rm
Im}\,q_{n\downarrow}|a}$, we get for any positions of $\lambda$
and $\eta$ \beq{A3} \begin{array}{l} I(\lambda,\eta)=
\Frac{2\sin(\lambda-\eta)a/2} {\lambda-\eta} \left({\cal
F}(\lambda)+{\cal F}(\eta)\right) \\[\bigskipamount]
\phantom{I(\lambda,\eta)}+ \Frac{2\cos(\lambda-\eta)a/2}
{\lambda-\eta} \left({\cal P}(\lambda)-{\cal P}(\eta)\right),
\end{array} \eeq where the function ${\cal P}(\lambda)$ is defined by
the residues in poles of ${\cal F}(q_{\perp})$ \beq{A4} {\cal
P}(\lambda)=-i\left( \sum_{n\uparrow} \frac{{\rm
res}_{n\uparrow}{\cal F}} {\lambda-q_{n\uparrow}}-
\sum_{n\downarrow} \frac{{\rm res}_{n\downarrow}{\cal F}}
{\lambda-q_{n\downarrow}}\right). \eeq

To apply the results for (\ref{5.5}), one needs to specify the pole
structure of the integrand. For
the factors $F_{l,t}(q_{\perp})$, it is evident from (\ref{5.5.2})
and (\ref{5.5.3}). The
function $\bar{\Omega}_{l,t}(q_{\perp})$ is given by (\ref{3n.19.2}) and can be
presented as the sum of four simple pole terms. The poles are located
symmetrically with respect to the axes at the points
$x\pm iy$ and $-x\pm iy$, where
\beq{5.7}
x=\sqrt{\frac{\sqrt{p^4+\Gamma^4}-p^2}{2}},\quad
y=\sqrt{\frac{\sqrt{p^4+\Gamma^4}+p^2}{2}}
\eeq
are real and positive. The function $\bar\Omega(q_{\perp})$ takes the form
\beq{5.8}
\bar\Omega(q_{\perp})= \frac{1}{4i|z|^2}
\left(U(q_{\perp},z^*)-U(q_{\perp},z)\right),
\eeq
where $z=x+iy$ and
\beq{5.9} U(\alpha,z)= \frac{z}{\alpha-z^*}+
\frac{z^*}{\alpha+z}.
\eeq
This complex function of two complex variables will be extensively
used below and it is worthwhile to exhibit its symmetry
\beq{5.9.2}
U(\alpha,z)= U(-\alpha,-z)=
-U(-\alpha,z^*)= U^*(\alpha^*,z^*).
\eeq
The real and imaginary parts of this function for the first argument
$\beta+i\gamma$ are of the form
\beq{8.12}
{\rm Re}\,U(\beta+i\gamma,z)=
\frac{2\beta x}{\Delta}\left(\beta^2+
(\gamma+2y)^2-|z|^2\right),
\eeq
\beq{8.11}
{\rm Im}\,U(\beta+i\gamma,z)=
-\frac{2x}{\Delta}
\left(\beta^2\gamma+(\gamma+2y)(x^2+(\gamma+y)^2)\right),
\eeq
where
\beq{8.11.2}
\Delta=\left(\beta^2+x^2+(\gamma+y)^2\right)^2-4\beta^2x^2.
\eeq
Note, that $U(i\gamma,z)$ is pure imaginary and given by
\beq{6m.9.2}
{\rm Im}\,U(i\gamma,z)=-\frac{2x(\gamma+2y)}{x^2+(\gamma+y)^2}.
\eeq

Now, we are ready to apply the general result (\ref{A3}) and
(\ref{A4}) to the integral (\ref{5.5}). It is enough to find three
"elementary"
integrals $I_n(\lambda,\eta)$ ($n=1,2,3$) with \beq{A5}
\begin{array}{l} {\cal F}_1(q_{\perp})=\bar\Omega(q_{\perp}),
\\[\bigskipamount]
{\cal F}_2(q_{\perp})=
\Frac{\bar\Omega(q_{\perp})}{q_{\|}^2+q_{\perp}^2},
\\[\bigskipamount]
{\cal F}_3(q_{\perp})= \Frac{q_{\perp}\bar\Omega(q_{\perp})}
{q_{\|}^2+q_{\perp}^2}. \end{array} \eeq All ${\cal
F}_n(q_{\perp})$ have the poles from $\bar\Omega(q_{\perp})$, the last
two have the additional poles in $q_{\perp}=\pm iq_{\|}$. The
functions ${\cal P}_n(\lambda)$ take the form \beq{A6}
\begin{array}{l} {\cal P}_1(\lambda)= -\Frac{1}{4|z|^2}
\left(U(\lambda,z^*)+U(\lambda,z)\right),
\\[\bigskipamount]
{\cal P}_2(\lambda)= \Frac{{\cal
P}_1(\lambda)}{q_{\|}^2+\lambda^2}- \Frac{i\lambda
U(iq_{\|},z)}{2q_{\|}|z|^2(q_{\|}^2+\lambda^2)},
\\[\bigskipamount]
{\cal P}_3(\lambda)= \Frac{{\lambda\cal
P}_1(\lambda)}{q_{\|}^2+\lambda^2}+
\Frac{iq_{\|}U(iq_{\|},z)}{2|z|^2(q_{\|}^2+\lambda^2)}.
\end{array}
\eeq Note, that ${\cal F}_n$ and ${\cal P}_n$, which determine
$I_n(\lambda,\eta)$ by (\ref{A3}), contain the function $U(\alpha,z)$
(\ref{5.9}) varied by first argument.

The function $J(\lambda,\eta)$ (\ref{5.5})
can be presented as a sum of
$I_n(\lambda,\eta)$ with coefficients evident from (\ref{5.5.2})
and (\ref{5.5.3}). After some rearrangement the results can be
written in the form
\beq{5.10}
J_{l,t}(\lambda,\eta)=J^{(1)}_{l,t}(\lambda,\eta)+
J^{(2)}_{l,t}(\lambda,\eta). \eeq Both terms
in the right-hand side are proportional to
the factors $U(\alpha,\beta)$, however, $J^{(1)}$ includes all
terms with $\alpha=\lambda$ or $\eta$, while $J^{(2)}$ includes
terms with $\alpha=iq_{\|}$. Namely,
\beq{5.11}
J^{(1)}_{l,t}(\lambda,\eta)=
\frac{e^{-i\efrac{(\lambda-\eta)a}{2}}
\Lambda_{l,t}(\lambda,\eta)-e^{i\efrac{(\lambda-\eta)a}{2}}
\Lambda_{l,t}(\eta,\lambda)}{\lambda-\eta},
\eeq
\beq{5.12}
J^{(2)}_{l,t}(\lambda,\eta)=
\cos\frac{(\lambda-\eta)a}{2}\,
\Phi_{l,t}(\lambda,\eta)\,
\frac{{\rm Im}\,U(iq_{\|},z)}{q_{\|}|z|^2},
\eeq
where
\beq{5.13}
\Lambda_{l,t}(\lambda,\eta)=
-\Frac{F_{l,t}(\lambda)U(\lambda,z)-
F_{l,t}(\eta)U(\eta,z^*)}{2|z|^2},
\eeq
\beq{5.16}
\begin{array}{l}
\Phi_{l,t}(\lambda,\eta)=
\pm\Frac{1}{(q_{\|}^2+\lambda^2)(q_{\|}^2+\eta^2)}\times{}
\\[\bigskipamount]
\phantom{\Phi_{l,t}(\lambda)}\times
\left( q_{\|}^2({\bf Q}_{\|}{\bf q}_{\|}+
\lambda Q^{\sigma\sigma'}_{\perp})
({\bf Q}_{\|}{\bf q}_{\|}+
\eta {Q^{\tau\tau'}_{\perp}}^*)-{}\right.
\\[\bigskipamount]
\phantom{\Phi_{l,t}(\lambda)\times}-
\left.(\lambda{\bf Q}_{\|}{\bf q}_{\|}-
Q^{\sigma\sigma'}_{\perp}q_{\|}^2)
(\eta{\bf Q}_{\|}{\bf q}_{\|}-
{Q^{\tau\tau'}_{\perp}}^*q_{\|}^2)\right),
\end{array}
\eeq
and the functions $F_{l,t}$ are given by
(\ref{5.5.2}) and (\ref{5.5.3}).
In (\ref{5.16}) "+" and "-" correspond to "$l$" and
"$t$", respectively.

Note, that the dependence on the parameter $a$, the thickness of the
layer, is explicitly indicated in (\ref{5.11}) and
(\ref{5.12}). All other functions are independent on $a$. The
functions $F$ and $\Phi$ are
constructed entirely from kinematic factors, initial and final
neutron momenta. Parameters of correlation function, $z=x+iy$,
apart from the explicit factor $|z|^2$, enter only in the function
$U(\alpha,z)$ with $\alpha=\lambda,\eta$ or $iq_{\|}$.

The cross section (\ref{5.2}), (\ref{5.3}) is the function of the
incoming and outcoming neutron wave vectors, which are included in
parameters ${\bf Q}_{\|}={\bf k}_{\|}-{\bf k}'_{\|}$ (\ref{3.8.2})
and $Q^{\sigma\sigma'}_{\perp}$, $Q^{\tau\tau'}_{\perp}$
(\ref{3.8.4}). For the crystal
these parameters are split into the parts inside the first
Brillouin zone, ${\bf q}_{\|}$, $\lambda$, $\eta$, and
corresponding reciprocal lattice vectors ${\bf G}_{\|}$,
$G_{\lambda}$, $G_{\eta}$. Note, that
\beq{5.17}
F_l(\lambda)=
F_l(\eta)= Q_{\|}^2+
Q^{\sigma\sigma'}_{\perp}{Q^{\tau\tau'}_{\perp}}^*,
\eeq
\beq{5.17.2}
F_t(\lambda)=F_t(\eta)=0,\quad
\Phi_{l,t}(\lambda,\eta)=\pm Q_{\|}^2,
\eeq
when all ${\bf G}_{\|}$, $G_{\lambda}$, $G_{\eta}$ are
zero.

Thus, with calculated $J(\lambda,\eta)$, the general expression for
the inelastic cross section is of the form
(\ref{5.2}), (\ref{5.3}) as the sum over the indices $\sigma$,
$\sigma'$, $\tau$, $\tau'$. This summation, for a general case,
is cumbersome because the variables $\lambda$ and
$\eta$ (\ref{5.4}) enter the functions $U(\lambda,z)$ and
$U(\eta,z)$ in a sophisticated way.

\subsection{Inelastic cross section for initial sub-barrier neutrons
\label{Inelastic cross section for initial sub-barrier neutrons}}

In this part of Appendix we specify the general expression for the
inelastic cross section (\ref{5.2}), (\ref{5.3}) for the case,
when the initial neutron is sub-barrier.
The final energy can be both below and
above the barrier. The scheme of approximation varies for these
two cases, and they will be considered separately. The object of
analysis will be the factor $\Pi_{l,t}$ in the cross section
(\ref{5.2}). All other factors are universal and will be added
afterwards.

For the initial sub-barrier neutron it is useful to define
\beq{6m.1}
\bar k=i\kp,\quad
\kp=\sqrt{k^2_0-k^2_{\perp}}.
\eeq
The quantity $\gamma=e^{-\kp a}\to 0$ for the initial
channel may be neglected  everywhere.
In particular, for the neutron coming
from the left we have
\beq{6m.3}
A^{(L)}_{+1}=1,\quad
A^{(L)}_{-1}=-r\gamma\to 0.
\eeq

To compensate
$\gamma$ in the first factor in (\ref{5.3}) we need
from the sum $\gamma^{-1}=e^{\kp a}$.  Therefore, the factor needed
comes only from the first exponent in (\ref{5.11}) with
$\sigma=\tau=1$.
The same exponent should be taken into account in
(\ref{5.12}).

Following (\ref{5.10}), the $\Pi$ factor (\ref{5.3}) can be
presented as the sum of two terms
\beq{6m.4}
\begin{array}{l}
\Pi^{(1)}_{l,t}=
\Frac{|\gamma'|}{|1-r'^2\gamma'^2|^2}\,
{\ds\sum_{\sigma',\tau'}}
A'_{\sigma'}A'^*_{\tau'}\times{}
\\[\bigskipamount]
\phantom{\Pi^{(1)}_{l,t}}\times{}
(\gamma')^{-\sigma'/2}(\gamma'^*)^{-\tau'/2}\,
\Frac{\Lambda^{\sigma'\tau'}_{l,t}(\lambda,\eta)}
{\lambda-\eta},
\end{array}
\eeq
\beq{6m.5}
\begin{array}{l}
\Pi^{(2)}_{l,t}=
\Frac{|\gamma'|}{|1-r'^2\gamma'^2|^2}\,
\Frac{{\rm Im}\,U(iq_{\|},z)}{2q_{\|}|z|^2}
{\ds\sum_{\sigma',\tau'}}
A'_{\sigma'}A'^*_{\tau'}\times{}
\\[\bigskipamount]
\phantom{\Pi^{(2)}_{l,t}}\times
(\gamma')^{-\sigma'/2}(\gamma'^*)^{-\tau'/2}\,
\Phi^{\sigma'\tau'}_{l,t}(\lambda,\eta).
\end{array}
\eeq
Parameters $\lambda$
and $\eta$ are defined by (\ref{5.4}) and can be rewritten now
as
\beq{6m.6}
\lambda =i\kp+\sigma'\bar k'_1,\quad
\eta=-i\kp+\tau'\bar k'^*_1,
\eeq
where $\bar k'_1=\bar k'-G_{\perp}$ is the
part of $\bar k'$ in the first Brillouin zone
($G_{\perp}=\sigma'G_{\lambda}=\tau'G_{\eta}$).
The identity $e^{iG_{\perp}a}=1$
for the reciprocal lattice vector is taken
into account.
The functions $\Lambda$ and $\Phi$ in (\ref{6m.4})
and (\ref{6m.5}) depend on indices
$\sigma'$ and $\tau'$ via their arguments.

\subsubsection{Final state: below the barrier}

To be more precise, only normal to the
surface component of the final neutron energy should be below the
barrier. For the normal component of the final neutron wave vector
inside the sample we introduce (with analogy to (\ref{6m.1}))
\beq{6m.7}
\bar k'=i\kp',\quad \kp'=\sqrt{k^2_0-k'^2_{\perp}}.
\eeq
In the sums (\ref{6m.4}), (\ref{6m.5}) over $\sigma'$ and $\tau'$
the main term corresponds evidently to $\sigma'=\tau'=1$ because
the others contain the factor $\gamma'=e^{-\kp' a}$, which may be
neglected apart from a small vicinity of the barrier, where
$k'_{\perp}\to k_0$, $\kp' a\leq 1$ and $\gamma'$ may approach
the unity. The barrier region is discussed in details later. Note that
even for $\sigma'=\tau'=1$ only the coefficients $A'^{(L)}$
contribute to (\ref{6m.4}), (\ref{6m.5}), but not $A'^{(R)}$.
Surely, it means that the neutrons coming from the left are
scattered also to the left.

For the small energy transfer one may put ${\bf G}_{\|}=0$ and
$G_{\lambda}=G_{\eta}=0$. Taking into account (\ref{5.17})
and (\ref{5.17.2}), we get
\beq{6m.8}
\begin{array}{l}
\Pi^{(1)}_l=
\Frac{\Lambda^{11}_l(\lambda_0,\lambda_0^*)}{2\lambda_0}=
-\Frac{(Q_{\|}^2+|\lambda_0|^2)\,{\rm Im}\,U(\lambda_0,z)}
{2|\lambda_0||z|^2},
\\[\bigskipamount]
\Pi^{(1)}_t=0,
\end{array}
\eeq
\beq{6m.9}
\Pi^{(2)}_{l,t}=
\Phi^{11}_{l,t}(\lambda_0,\lambda_0^*)\,
\frac{{\rm Im}\,U(iQ_{\|},z)}{2Q_{\|}|z|^2}=
\pm\frac{Q_{\|}\,{\rm Im}\,U(iQ_{\|},z)}{2|z|^2},
\eeq
where
\beq{6m.9.3}
\lambda_0=i(\kp+\kp').
\eeq
Note, that for pure imaginary first argument $\alpha=i\gamma$ the
function ${\rm Im}\,U$ is given by (\ref{6m.9.2}).

\subsubsection{Final state: above the barrier}

In the sums (\ref{6m.4}), (\ref{6m.5}) over $\sigma'$ and $\tau'$
all terms should be considered equally important, since
$\gamma'=e^{i\bar k'a}$ is now not a small parameter. If we
neglect radiative capture, then the quantities $\bar k'$ and $r'$
are real, and $|\gamma'|=1$.

It is useful to split (\ref{6m.4}), (\ref{6m.5}) into the sums
with $\sigma'=\tau'$ (where $\eta=\lambda^*$) and that with
$\sigma'=-\tau'$ (where $\eta=-\lambda$)
\beq{6m.14}
\begin{array}{l}
\Pi^{(1)}_{l,t}= \Frac{1}{|1-r'^2\gamma'^2|^2}\left(
{\ds\sum_{\sigma'}}|A'_{\sigma'}|^2
\Frac{\Lambda^{\sigma'\sigma'}_{l,t}(\lambda,\lambda^*)}
{2i\kp}+{}\right.
\\[\bigskipamount]
\phantom{\Pi^{(1)}_{l,t}= \Frac{1}{|1|^2}}\left.+
{\ds\sum_{\sigma'}} A'_{\sigma'}A'^*_{-\sigma'}
(\gamma')^{-\sigma'}
\Frac{\Lambda^{\sigma'\,-\sigma'}_{l,t}(\lambda,-\lambda)}
{2\lambda}\right),
\end{array}
\eeq
\beq{6m.15}
\begin{array}{l}
\Pi^{(2)}_{l,t}=
\Frac{{\rm Im}\,U(iq_{\|},z)}
{2q_{\|}|z|^2|1-r'^2\gamma'^2|^2}\,
\left( {\ds\sum_{\sigma'}}
|A'_{\sigma'}|^2
\Phi^{\sigma'\sigma'}_{l,t}(\lambda,\lambda^*)+{}\right.
\\[\bigskipamount]
\phantom{\Pi^{(2)}_{l,t}= \Frac{1}{|1|^2}}\left. +
{\ds\sum_{\sigma'}}A'_{\sigma'}A'^*_{-\sigma'}
(\gamma')^{-\sigma'}
\Phi^{\sigma'\,-\sigma'}_{l,t}(\lambda,-\lambda)\right).
\end{array} \eeq

For the explicit summation in (\ref{6m.14}) we need to know how
$\Lambda^{\sigma'\sigma'}_{l,t}(\lambda,\lambda^*)$ and
$\Lambda^{\sigma'\,-\sigma'}_{l,t}(\lambda,-\lambda)$
depend on $\sigma'$. From
(\ref{5.13}) with the use of symmetry properties (\ref{5.9.2}) one
can show that the quantity
\beq{6m.16}
\Lambda^{11}_{l,t}(\lambda,\lambda^*)\equiv
\Lambda^{11}_{l,t}(\lambda_0,\lambda_0^*)
\eeq
is pure imaginary,
where
\beq{6m.17}
\lambda_0=i\kp+\bar k'_1.
\eeq
On the other hand
\beq{6m.18}
\Lambda^{-1\,-\!1}_{l,t}(\lambda,\lambda^*)\equiv
\Lambda^{-1\,-\!1}_{l,t}(-\lambda^*_0,-\lambda_0)=
\Lambda^{11}_{l,t}(\lambda_0,\lambda_0^*),
\eeq
and
\beq{6m.19}
\begin{array}{l}
\Lambda^{1\,-\!1}_{l,t}(\lambda,-\lambda)\equiv
\Lambda^{1\,-\!1}_{l,t}(\lambda_0,-\lambda_0),
\\[\bigskipamount]
\Lambda^{-1\,1}_{l,t}(\lambda,-\lambda)\equiv
\Lambda^{-1\,1}_{l,t}(-\lambda^*_0,\lambda^*_0)=
-{\Lambda_{l,t}^{1\,-\!1}\,}^*(\lambda_0,-\lambda_0).
\end{array}
\eeq
With the
help of (\ref{6m.16})--(\ref{6m.19}) one readily finds
\beq{6m.20}
\begin{array}{l}
\Pi^{(1)}_{l,t}= \Frac{1}{|1-r'^2\gamma'^2|^2}\left(
(1+r'^2)\Frac{{\rm Im}\,\Lambda^{11}_{l,t}(\lambda_0,\lambda^*_0)}
{2\kp}+{}\right.
\\[\bigskipamount]
\phantom{\Pi^{(1)}_{l,t}}\left.+ 2{\rm Re}
\left(A'_{+1}A'^*_{-1}(\gamma')^{-1}
\Frac{\Lambda^{1\,-\!1}_{l,t}(\lambda_0,-\lambda_0)}
{2\lambda_0}\right)\right).
\end{array}
\eeq

The sum (\ref{6m.15}) in a similar way results in
\beq{6m.21}
\begin{array}{l}
\Pi^{(2)}_{l,t}=
\Frac{{\rm Im}\,U(iq_{\|},z)}
{2q_{\|}|z|^2|1-r'^2\gamma'^2|^2}\,
\left((1+r'^2)\Phi^{11}_{l,t}(\lambda_0,\lambda_0^*)+{}\right.
\\[\bigskipamount]
\phantom{\Pi^{(2)}_{l,t}}\left.+ 2{\rm
Re}\left(A'_{+1}A'^*_{-1}(\gamma')^{-1}
\Phi^{1\,-\!1}_{l,t}(\lambda_0,-\lambda_0)\right)\right),
\end{array}
\eeq
where the quantity $\Phi^{11}_{l,t}(\lambda_0,\lambda_0^*)$
(\ref{5.16}) is real.

The last terms in (\ref{6m.20}) and (\ref{6m.21}) have different
values for backward (B) and forward (F) scattering because
\beq{6m.22} \left[A'_{+1}A'^*_{-1}(\gamma')^{-1}\right]^B=
A'^{(L)}_{+1}A'^{(L)*}_{-1}(\gamma')^{-1}= -r'(\gamma')^{-2}, \eeq
\beq{6m.23} \left[A'_{+1}A'^*_{-1}(\gamma')^{-1}\right]^F=
A'^{(R)}_{+1}A'^{(R)*}_{-1}(\gamma')^{-1}=-r'. \eeq

\subsubsection{Final state: near the barrier. Resonances of penetration}

With the results of two previous subsections we are able to consider
the upscattering of UCN into the regions close to the barrier, below or
above.

In the low vicinity of the barrier $\kp'\ll\kp$. Above the barrier
two "vicinities" should be distinguished, "wide vicinity",
where $\bar k'\ll k_{\perp}$, but $\bar k'a\gg 1$, and "close
vicinity", where $\bar k'a\leq1$.

In the "wide vicinity" of the barrier, $\kp'$ and $\bar k'$ may
be treated as small parameters.  Due to small energy transfer, we
take $G_{\lambda}=G_{\eta}=0$. In the approximation
$\lambda\simeq\lambda_0=i\kp$ and $\eta\simeq\lambda^*_0$ the
equations (\ref{6m.4}) and (\ref{6m.5}) can be written as
\beq{6m.24}
\Pi^{(1)}_{l,t}\simeq
\Xi(r',\gamma')\,
\frac{\Lambda^{11}_{l,t}(\lambda_0,\lambda_0^*)}{2i\kp},
\eeq
\beq{6m.24.2}
\Pi^{(2)}_{l,t}\simeq \Xi(r',\gamma')
\Phi^{11}_{l,t}(\lambda_0,\lambda_0^*)\,
\frac{{\rm Im}\,U(iQ_{\|},z)}{2Q_{\|}|z|^2},
\eeq
where the factor
\beq{6m.25}
\Xi(r',\gamma')=
\frac{|\gamma'|}{|1-r'^2\gamma'^2|^2}\, \sum_{\sigma',\tau'}
A'_{\sigma'}A'^*_{\tau'}\,
(\gamma')^{-\sigma'/2}(\gamma'^*)^{-\tau'/2}
\eeq
for backward and
forward scattering is equal to
\beq{6m.26}
\Xi^B(r',\gamma')=
\frac{|1-r'\gamma'^2|^2}{|1-r'^2\gamma'^2|^2},\quad
\Xi^F(r',\gamma')=
\frac{|\gamma'|^2\,|1-r'|^2}{|1-r'^2\gamma'^2|^2}.
\eeq

When the final momentum $\bar k'$ is just above the barrier, where
it is reasonable to put $r'\simeq 1$, then $\Xi^B\to 1$ and
$\Xi^F\to 0$, and the corresponding barrier values of $\Pi _{l,t}$
(\ref{6m.24}) and (\ref{6m.24.2}) coincide with that from (\ref{6m.8}) and
(\ref{6m.9}). The deviations arise only close to the points where
\beq{6m.27}
\gamma'^2=e^{2i\bar k'a}=1 \quad\Longrightarrow\quad
\bar k'a=\pi n,\quad n=1,2,3\ldots
\eeq
and, therefore, a difference between $r'$ and 1 is of importance. For
these energies $\Pi^B$
decreases and $\Pi^F$ increases up to the peak values
\beq{6m.28}
\Pi^{(1)B,F}_{l,t}\simeq
\frac{1}{4}\,
\frac{\Lambda^{11}_{l,t}(\lambda_0,\lambda_0^*)}{2i\kp},
\eeq
\beq{6m.28.2}
\Pi^{(2)B,F}_{l,t}\simeq
\frac{1}{4}\,
\Phi^{11}_{l,t}(\lambda_0,\lambda_0^*)\,
\frac{{\rm Im}\,U(iQ_{\|},z)}{2Q_{\|}|z|^2}.
\eeq
It is seen from
(\ref{6m.27}) that these minima in $\Pi^B$ and maxima in $\Pi^F$
arise when the thickness of the layer is divisible by the half of
the neutron wave length in the substance. Thus, the reason for the
oscillation of the cross section is the resonances of penetration
through the layer.

When the final momentum is not too close to the barrier, the
factors containing $\gamma'=e^{i\bar k'a}$ (e.g.
$\left|1-r'^2\gamma'^2\right|^{-2}$) strongly oscillate. With
increase of the neutron final energy the amplitude of oscillations
goes down because $r'\to 0$. However, in the ''wide vicinity'' of
the barrier the oscillations are of importance. For not very thin
sample these oscillations cannot be resolved and only the smoothed
cross section is of interest.

\subsubsection{Smoothing over oscillations}

The functions $\Pi$ (\ref{6m.20}) and (\ref{6m.21}) with account
for (\ref{6m.22}) and (\ref{6m.23}) oscillate due to the factors
\beq{6m.29} \begin{array}{l} \Frac{1}{|1-r'^2\gamma'^2|^2}=
{\ds\sum_{l,m=0}^{\infty}} r'^{2(l+m)}\, e^{2i(l-m)\bar k'a},
\\[\bigskipamount]
\Frac{(\gamma')^{-2}}{|1-r'^2\gamma'^2|^2}=
{\ds\sum_{l,m=0}^{\infty}} r'^{2(l+m)}\, e^{2i(l-m-1)\bar k'a}.
\end{array}
\eeq
Keeping only the terms with zero indices in the exponents, we get
the smoothed functions
\beq{6m.30}
\begin{array}{l}
\av{\Pi^{(1)B,F}_{l,t}}=
\Frac{1}{1-r'^4}\left( (1+r'^2)
\Frac{{\rm Im}\,\Lambda^{11}_{l,t}(\lambda_0,\lambda^*_0)}
{2\kp}-{}\right.
\\[\bigskipamount]
\phantom{\av{\Pi^{(1)B,F}_{l,t}}= \Frac{1}{1-r'^4}}\left. -
2r'^n\,{\rm Re}\,
\Frac{\Lambda^{1\,-\!1}_{l,t}(\lambda_0,-\lambda_0)}
{2\lambda_0}\right),
\end{array}
\eeq
\beq{6m.31}
\begin{array}{l}
\av{\Pi^{(2)B,F}_{l,t}}=
\Frac{{\rm Im}\,U(iq_{\|},z)}
{2q_{\|}|z|^2(1-r'^4)}
\left((1+r'^2)\Phi^{11}_{l,t}(\lambda_0,\lambda_0^*)-{}
\right.
\\[\bigskipamount]
\phantom{\av{\Pi^{(2)B,F}_{l,t}}=
\Frac{{\rm Im}\,U(iq_{\|},z)}
{2q_{\|}|z|^2}}-\left.
2r'^n\,{\rm Re}\,
\Phi^{1\,-\!1}_{l,t}(\lambda_0,-\lambda_0)\right),
\end{array}
\eeq
where $n=3$ for backward and $n=1$ for forward
scattering, and $\lambda_0$ is given by (\ref{6m.17}).

The smoothing is not justified in the "close vicinity" of the
barrier where $\bar k'a\sim 1$. But since the smoothed quantities
(\ref{6m.30}) and (\ref{6m.31}) have the same limiting values as
(\ref{6m.24}) and (\ref{6m.24.2}), one may use the smoothed cross
section everywhere above the barrier.

The functions $\Lambda_{l,t}$, which enter (\ref{6m.30}),
are disclosed from (\ref{5.13}), (\ref{5.5.2}) and (\ref{5.5.3}) as
\beq{6m.132}
\begin{array}{l}
{\rm Im}\,\Lambda_l^{11}(\lambda_0,\lambda_0^*)={}
\\[\bigskipamount]
\phantom{}\!=\!
-\,{\rm Im}\left(\Frac{({\bf Q}_{\|}{\bf q}_{\|}+Q^{11}_{\perp}\lambda_0)
({\bf Q}_{\|}{\bf q}_{\|}+{Q^{11\,}_{\perp}}^*\lambda_0)\,
U(\lambda_0,z)}
{|z|^2(q^2_{\|}+\lambda_0^2)}\right),
\end{array}
\eeq
\beq{6m.133}
\begin{array}{l}
{\rm Im}\,\Lambda_t^{11}(\lambda_0,\lambda_0^*)={}
-\,{\rm Im}\left(\left(Q^2_{\|}+|Q_{\perp}^{11}|^2-{}
\phantom{\lefteqn{\Frac{{\bf Q}_{\|}{\bf q}_{\|}}
{q^2_{\|}+\lambda_0^2}}}
\right.\right.
\\[\bigskipamount]
\left.\left.\phantom{}-
\Frac{({\bf Q}_{\|}{\bf q}_{\|}+Q^{11}_{\perp}\lambda_0)
({\bf Q}_{\|}{\bf q}_{\|}+{Q^{11\,}_{\perp}}^*\lambda_0)}
{q^2_{\|}+\lambda_0^2}\right)
\Frac{U(\lambda_0,z)}{|z|^2}\right),
\end{array}
\eeq
\beq{6m.134}
\Lambda_l^{1\,-\!1}(\lambda_0,-\lambda_0)=
-\frac{\left(({\bf Q}_{\|}{\bf q}_{\|})^2-
(Q_{\perp}^{11}\lambda_0)^2\right)U(\lambda_0,z)}
{|z|^2(q_{\|}^2+\lambda_0^2)},
\eeq
\beq{6m.135}
\begin{array}{l}
\Lambda_t^{1\,-\!1}(\lambda_0,-\lambda_0)=
-\left(Q^2_{\|}-(Q_{\perp}^{11})^2-
\phantom{\lefteqn{\Frac{({\bf Q}_{\|}{\bf q}_{\|})^2}
{q_{\|}^2+\lambda_0^2}}}\right.
\\[\bigskipamount]
\left.\phantom{\Lambda_t^{1\,-\!1}(\lambda_0,-\lambda_0)}-
\Frac{({\bf Q}_{\|}{\bf q}_{\|})^2-
(Q_{\perp}^{11}\lambda_0)^2}
{q_{\|}^2+\lambda_0^2}\right)
\Frac{U(\lambda_0,z)}{|z|^2},
\end{array}
\eeq
with
\beq{6m.136}
Q_{\perp}^{11}=i\kp+\bar k',\quad
\lambda_0=i\kp+\bar k'_1.
\eeq
The functions $\Phi_{l,t}$, which enter (\ref{6m.31}), are evident
from (\ref{5.16}).

\subsubsection{Interim conclusion}

The general formula for the inelastic cross section (\ref{5.2})
for the initial sub-barrier neutrons is now fully
defined with $\Pi_{l,t}=\Pi_{l,t}^{(1)}+\Pi_{l,t}^{(2)}$ given by
(\ref{6m.8}) and (\ref{6m.9}) for the final state below the
barrier ($k'_{\perp}<k_0$) or by (\ref{6m.30}) and (\ref{6m.31}) for
that above the barrier ($k'_{\perp}>k_0$).

\subsection{Inelastic transition probability for ${\bf G}=0$}

In this part of Appendix we calculate the dimensionless factor (\ref{7n.5}),
which determines the inelastic transition probability (\ref{7n.4})
with momentum transfer inside the first Brillouin zone (${\bf G}=0$).

\begin{figure}
\begin{center}
\mbox{\includegraphics*[scale=0.5]{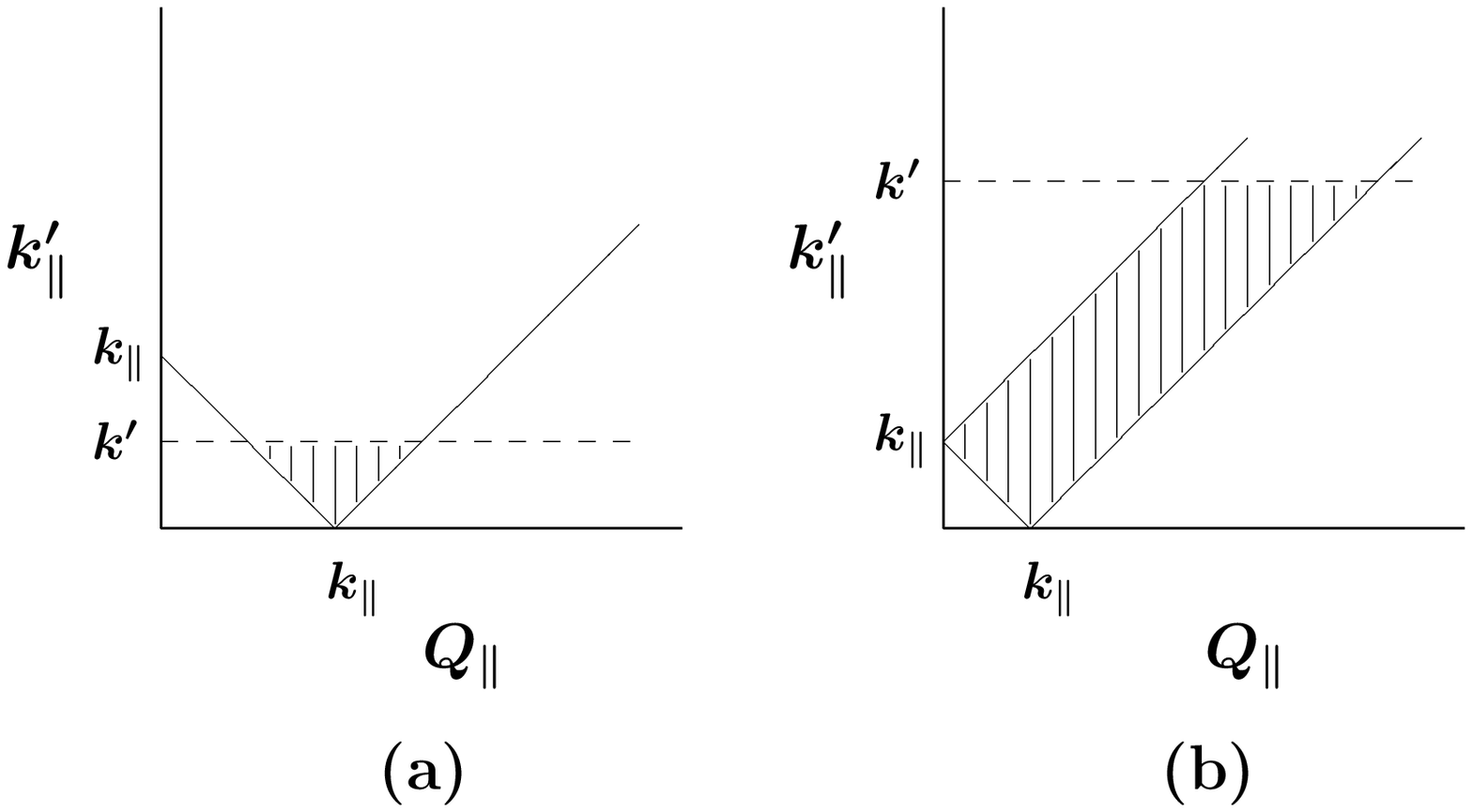}}
\caption{\label{F7} Integration area for the double integral (\ref{8.7});
the cases $k' < k_{\|}$ (a) and $k' > k_{\|}$ (b).}
\end{center}
\end{figure}

In the integral (\ref{7n.5})
it is useful to replace the variable $k'_{\perp}$ by $k'^2_{\|}$ and
the variable $\varphi$ by $Q^2_{\|}=({\bf k}'_{\|}-{\bf k}_{\|})^2$
running
from $(k'_{\|}-k_{\|})^2$ up to $(k'_{\|}+k_{\|})^2$. Thus,
\beq{8.7}
\begin{array}{l}
W_{l,t}(k_{\perp},k_{\|}\to\ve')={}
\\[\bigskipamount]
\phantom{W_{l,t}}=
\Frac{k_{\perp}}{2\pi^2k}
{\ds\int} dk'^2_{\|}{\ds\int} dQ^2_{\|}\,
g(Q_{\|},k'_{\|},k_{\|})\,
\Frac{|t'|^2}{k'_{\perp}}\,
\Pi_{l,t},
\end{array}
\eeq
where the function $g$ can be presented in two forms
\beq{8.8}
\begin{array}{l}
g(Q_{\|},k'_{\|},k_{\|})=
\Frac{1}{\sqrt{\left(\!(k'_{\|}\!+\!k_{\|})^2\!-\!Q^2_{\|}\right)\!
\left(Q^2_{\|}\!-\!(k'_{\|}\!-\!k_{\|})^2\!\right)}}={}
\\[\bigskipamount]
\phantom{W(Q_{\|})}=
\Frac{1}{\sqrt{\left(\!(Q_{\|}\!+\!k_{\|})^2\!-\!k'^2_{\|}\right)\!
\left(k'^2_{\|}\!-\!(Q_{\|}\!-\!k_{\|})^2\!\right)}}\,.
\end{array}
\eeq
Note, that
\beq{7n.5.2}
|t'|^2=\left\{\begin{array}{ll}
4k'^2_{\perp}/k^2_0, & k'_{\perp}<k_0,\\
4k'^2_{\perp}/(k'_{\perp}+\bar k')^2, & k'_{\perp}>k_0.
\end{array}\right.
\eeq
Integration area for double integral (\ref{8.7}) is shown in Fig.~\ref{F7}.

The general formulae for the factors $\Pi_{l,t}$ are given by
(\ref{6m.8}), (\ref{6m.9}) for the final state below the
barrier ($k'_{\perp}<k_0$), and by (\ref{6m.30}), (\ref{6m.31})
for the final state above the barrier ($k'_{\perp}>k_0$).
When ${\bf G}=0$, the Eqs.(\ref{6m.30}) and (\ref{6m.31})
are simplified, and the sums of forward and backward scattering
contributions can be written in the form
\beq{8.1}
\begin{array}{l}
\Pi_l^{(1)}=\Frac{1}{1-r'^2}\left(
-\Frac{(Q_{\|}^2+|\lambda_0|^2)\,{\rm Im}\,U(\lambda_0,z)}
{\kp |z|^2}+{}\right.
\\[\bigskipamount]
\left.\phantom{\Pi_l^{(1)}}+
\Frac{r'}{|z|^2}\,{\rm Re}\,\Frac{(Q_{\|}^2-\lambda_0^2)\,U(\lambda_0,z)}
{\lambda_0}\right),\quad
\Pi_t^{(1)}=0.
\end{array}
\eeq
\beq{8.2}
\Pi_{l,t}^{(2)}=\pm\frac{Q_{\|}\,{\rm Im}\,U(iQ_{\|},z)}
{(1+r')|z|^2},
\eeq
where $\lambda_0=i\kp+\bar k'$.

\subsubsection{Estimations for variables and functions
\label{sA.3.1}}

The function $U(\alpha,z)$ (\ref{5.9}) in (\ref{8.1}) and (\ref{8.2})
represents the correlation function. The real and imagine parts of the
variable $z=x+iy$ are determined by the equations
\beq{8.3}
2xy=\Gamma^2,\quad
y^2-x^2=Q_{\|}^2-\frac{\epsilon\,\omega^2}{c^2}.
\eeq

With the fixed initial and final neutron energies, $\ve$ and $\ve'$, the
"damping factor" $\Gamma^2$ (see Section~\ref{Correlation function})
may be treated as the constant (defined by appropriate physics). The
same is true for the quantity $\epsilon\,\omega^2/c^2$. It is
convenient to use one of these parameters as a unit and transform all
variables to the dimensionless form ($k'_{\|}\to\tilde k'_{\|}$,
$Q_{\|}\to\tilde Q_{\|}$, $\Pi_{l,t}\to\tilde\Pi_{l,t}$ etc.). We
choose as the unit $\Gamma/\sqrt{2}$, thus, e.g.,
\beq{8.4}
x\to \tilde x=\frac{\sqrt{2}x}{\Gamma},\quad
y\to \tilde y=\frac{\sqrt{2}y}{\Gamma},\quad
Q_{\|}\to \tilde Q_{\|}=\frac{\sqrt{2}Q_{\|}}{\Gamma},
\eeq
and, in particular,
\beq{8.5}
\tilde\omega^2\equiv\frac{2\epsilon\,\omega^2}{\Gamma^2c^2}.
\eeq
Note, that $\epsilon=1$, except for the longitudinal thermo-diffusion
mode (see text after (\ref{3n.19.2})), where $\epsilon=0$. In the
integral (\ref{8.7}) the upper limit for $\tilde k'^2_{\|}$ is
\beq{8.16}
\tilde k'^2=\frac{2k'^2}{\Gamma^2}.
\eeq

The equations (\ref{8.3}) take the form
\beq{8.6}
\tilde x\tilde y=1,\quad
\tilde y^2-\tilde x^2=\tilde Q_{\|}^2-\tilde\omega^2.
\eeq
They define the functions $\tilde x(\tilde Q_{\|})$ and $\tilde
y(\tilde Q_{\|})$ provided the parameter $\tilde\omega^2$ is given.
Therefore, for our problem (with fixed $\ve$ and $\ve'$), there is
only one additional parameter, $\tilde\omega^2$, which should be
specified by the physics of the process.

The damping factors $\Gamma^2$ for longitudinal sound waves both for
solids and liquids (\ref{3n.5}) and for transverse sound waves for
solids (\ref{3n.6}), as a rule, are much less than $k'^2$. On the
other hand, in the diffusion-like cases (\ref{3n.14.2}) and
(\ref{3n.18}) the damping factor takes the form
\beq{8.17}
\Gamma^2=\frac{|\omega|}{D}=
\frac{2m|\ve'-\ve|}{\hbar^2d},\quad
d=\frac{2mD}{\hbar},
\eeq
where $D=D_S$ for longitudinal thermo-diffusion both for solids and
liquids, and $D=\nu$ for damping transverse waves in liquids.
Parameter $d$ is dimensionless. It means, that (\ref{8.16}) can be
presented as
\beq{8.18}
\tilde k'^2=\frac{2\ve'd}{|\ve'-\ve|}.
\eeq
Note, that in all realistic cases $d\gg 1$ (see Section~\ref{s8}).
Therefore, both for sound-like and diffusion-like modes one has
\beq{8.19}
\tilde k'^2\gg 1.
\eeq

The following relationship
\beq{8.20}
\tilde\omega^2\ll\tilde k'^2,
\eeq
is also of great importance. Indeed,
\beq{8.21}
\frac{\omega^2}{c^2}\approx
\frac{\ve'^2}{\hbar^2c^2}=
\frac{\ve'}{2mc^2}\,k'^2\ll k'^2,
\eeq
because $\ve'$ is restricted from above by the Debye energy
\beq{8.22}
\ve_D=\hbar c(6\pi^2n)^{1/3},
\eeq
and $\ve_D\ll 2mc^2$ for all substances.

Let us separate two regions of integration oveq $\tilde Q_{\|}$ in
(\ref{8.7}): (I), where $\tilde Q_{\|}\sim\tilde k'\gg 1$ and
$\tilde\omega$ (see (\ref{8.19}) and (\ref{8.20})), therefore,
\beq{8.23}
{\rm (I)}\quad
\tilde x\ll 1\ll\tilde y\sim\tilde Q_{\|}\sim\tilde k',
\eeq
and (II), where $\tilde Q_{\|}\ll\tilde k'$, therefore (see
(\ref{8.19}) and (\ref{8.20})),
\beq{8.24}
{\rm (II)}\quad
\tilde x,\tilde y,\tilde Q_{\|}\ll\tilde k'.
\eeq

In the region (I) the factor $\Pi_{l,t}^{(2)}$ (\ref{6m.9}) and
(\ref{8.2}) is of the scale
\beq{8.25}
\tilde\Pi^{(2)}_{l,t}\sim
\frac{1}{\tilde y^3}\sim
\frac{1}{\tilde Q_{\|}^3}\ll 1.
\eeq
The explicit form of the factor $\tilde\Pi^{(1)}_l$ in the same region
depends on whether small or large energy transfers. For small the energy
transfer, $\tilde y\sim\tilde Q_{\|}\sim\tilde k'\sim\tilde{\kp}$,
and one has
\beq{8.26}
\tilde\Pi^{(1)}_l\sim
\frac{1}{\tilde Q_{\|}^3}\ll 1
\eeq
both from (\ref{6m.8}) and (\ref{8.1}). The case of the large energy
transfer corresponds to the relations: $\tilde{\kp}\sim\tilde
k_0\ll\tilde y\sim\tilde Q_{\|}\sim\tilde k'$. In this case only the
region $k'_{\perp}>k_0$ is of interest. Therefore, one obtains from
(\ref{8.1})
\beq{8.27}
\tilde\Pi^{(1)}_l\simeq
\frac{4\left(\tilde Q^2_{\|}+\tilde k'^2_{\perp}\right)}
{\tilde{\kp}\left(\tilde k'^2_{\perp}+\tilde y^2\right)^2}.
\eeq
Note, that this term dominates due to the relatively small factor
$\tilde{\kp}$ in its denominator.

Now let us consider the region (II). The dimensionless factor
$\tilde\Pi_{l,t}^{(2)}$ both below and above the barrier
may be written in the form
\beq{8.28}
\tilde\Pi^{(2)}_{l,t}=
\mp\frac{\phi'\tilde x\tilde Q_{\|}
(\tilde Q_{\|}+2\tilde y)}
{|\tilde z|^2
\left(\tilde x^2+(\tilde Q_{\|}+\tilde y)^2\right)},
\eeq
where
\beq{8.15}
\phi'=\left\{
\begin{array}{ll}
1, & k'_{\perp}<k_0,\\
(k'_{\perp}+\bar k')/k'_{\perp},\quad & k'_{\perp}>k_0.
\end{array}\right.
\eeq

The factor (\ref{6m.8}), (\ref{8.1}) for the small energy transfer takes the form
\beq{8.29}
\tilde\Pi^{(1)}_l\simeq
\frac{\phi'\tilde x}{|\tilde z|^2}.
\eeq
But it cannot be substantially simplified for large energy transfer.
However, this is of no importance. Indeed, in this case the main
contribution to the cross section comes from the region (I) due to
the dominating term (\ref{8.27}).

\subsubsection{Estimation of the integral}

Let us now estimate the integral (\ref{8.7}).
We see that the region (I) contributes only to the factor $W_l$ for
the large energy transfer due to (\ref{8.27}). Since $k'\sim
Q_{\|}\gg\kp\sim k_{\|}$, one can perform the integration
over $k'_{\|}$ in
(\ref{8.7}) in the narrow interval
$Q_{\|}-k_{\|}<k'_{\|}<Q_{\|}+k_{\|}$ (see Fig.~\ref{F7}b). It gives
\beq{8.30}
W^{(I)}_l(k_{\perp},k_{\|}\to\ve')\simeq
\frac{k_{\perp}}{2\pi k}
\int d\tilde Q^2_{\|}\,
\frac{|t'|^2}{\tilde k'_{\perp}}\,
\tilde\Pi^{(1)}_l,
\eeq
where $\tilde k'_{\perp}=\sqrt{\tilde k'^2-\tilde Q^2_{\|}}$, and
$\tilde\Pi^{(1)}_l$ is given by (\ref{8.27}). Taking into account
(\ref{8.6}) and (\ref{8.23}) and integrating over $\tilde Q^2_{\|}$
from $\tilde\omega^2$ to $\tilde k'^2$, one obtains
\beq{8.30.2}
W^{(I)}_l(k_{\perp},k_{\|}\to\ve')\simeq
\frac{2k_{\perp}\Gamma^2k'^2}
{\pi k\kp\left(k'^2-\epsilon\,\omega^2/c^2_l\right)^{3/2}}\,,
\eeq
where the result is presented in terms of the physical quantities.

Now let us consider the contribution to the cross section from the
region (II). It is convenient to rewrite the factor (\ref{8.7}) as
follows
\beq{8.31}
W_{l,t}(k_{\perp},k_{\|}\to\ve')=
\frac{k_{\perp}}{2\pi^2k}
\int\! d\tilde Q^2_{\|}
\frac{\tilde\Pi_{l,t}}{\phi'}\!
\int\! d\tilde k'^2_{\|}\,
\tilde g \tilde f(k'_{\perp}),
\eeq
where the function $\tilde\Pi_{l,t}/\phi'$ given by (\ref{8.28}) and
(\ref{8.29}) depends only on $\tilde Q_{\|}^2$, and
\beq{8.32}
f(k'_{\perp})\equiv
\Frac{\phi'|t'|^2}{k'_{\perp}}=
\Frac{4}{k^2_0}
\left(k'_{\perp}-
\theta(k'^2_{\perp}-k^2_0)
\sqrt{k'^2_{\perp}-k^2_0}\,\,\right),
\eeq
with $\theta(x)=0$ for $x<0$ and $\theta(x)=1$ for $x\ge 0$.

The integral over $\tilde k'^2_{\|}$ is of the form
\beq{8.33}
\int d\tilde k'^2_{\|}\,
\tilde g \tilde f(k'_{\perp})=
\frac{4}{\tilde k^2_0}
\left(I(\tilde k')\!-\!I(\sqrt{\tilde k'^2-\tilde k^2_0}\,)\right),
\eeq
where
\beq{8.34}
I(b)=\theta(b^2-\tilde k_{\|}^2)\int d\tilde k'^2_{\|}\,
\tilde g\,\,
\theta(b^2-\tilde k'^2_{\|})
\sqrt{b^2-\tilde k'^2_{\|}}.
\eeq
The variable $\tilde k'^2_{\|}$ runs from $(\tilde Q_{\|}-\tilde
k_{\|})^2$ up to $(\tilde Q_{\|}+\tilde k_{\|})^2$ (see Fig.~\ref{F7}).
Then, $I(b)$ can be presented in terms of elliptic integral
\beq{8.35}
I(b)=2\,\theta(b^2\!-\tilde k_{\|}^2)\,
\beta E(\frac{\alpha^2}{\beta^2}),\quad\!\!\!\!
E(m)\!=\!\!\int\limits_0^{\pi/2}\!\!\sqrt{1-m\sin^2\theta}\,d\theta,
\eeq
where
\beq{8.36}
\alpha^2=4\tilde Q_{\|}\tilde k_{\|},\quad
\beta^2=b^2-(\tilde Q_{\|}-\tilde k_{\|})^2.
\eeq

It is of importance, that in the region (II) both for small and large
energy transfer one has
\beq{8.37}
\tilde Q_{\|}\tilde k_{\|}\ll
\tilde k'\tilde k_{\|}\leq
\tilde k'^2-\tilde k^2_0,\,\tilde k'^2
\quad\Longrightarrow\quad
\alpha\ll\beta.
\eeq
Then, really $E(\alpha^2/\beta^2)\simeq\pi/2$, and, therefore, the
contribution from the region (II) to the cross section is
\beq{8.38}
W^{(I\!I)}_{l,t}(k_{\perp},k_{\|}\to\ve')\simeq
\frac{k_{\perp}}{2\pi k}
\int d\tilde Q^2_{\|}\,
\frac{|t'|^2}{\tilde k'_{\perp}}\,
\tilde\Pi_{l,t},
\eeq
where $\tilde k'_{\perp}=\sqrt{\tilde k'^2-(\tilde Q_{\|}-\tilde
k_{\|})^2}\simeq\sqrt{\tilde k'^2-\tilde k^2_{\|}}$, while
$\tilde\Pi^{(1)}_l$ and $\tilde\Pi^{(2)}_{l,t}$ are given by
(\ref{8.29}) and (\ref{8.28}), respectively.

To calculate the integral it is convenient to replace $\tilde
Q^2_{\|}$ by $\tilde x$ with the use of
\beq{8.39}
d\tilde Q^2_{\|}=-\frac{2|\tilde z|^2}{\tilde x}\,d\tilde x.
\eeq
The limit $\tilde Q_{\|}=0$ corresponds to
\beq{8.40}
\tilde x_0^2=\sqrt{\frac{\tilde\omega^4}{4}+1}+
\frac{\tilde\omega^2}{2}.
\eeq
In the opposite limit $\tilde Q_{\|}\to\infty$, $\tilde x\to 0$, then
the integral (\ref{8.38}) can be presented in the form
\beq{8.41}
W^{(I\!I)}_l(k_{\perp},k_{\|}\to\ve')=
\frac{k_{\perp}}{\pi k}\,
\tilde f(k'_{\perp})
\left(\tilde J^{(1)}-\tilde J^{(2)}\right),
\eeq
\beq{8.42}
W^{(I\!I)}_t(k_{\perp},k_{\|}\to\ve')=
\frac{k_{\perp}}{\pi k}\,
\tilde f(k'_{\perp})\,
\tilde J^{(2)},
\eeq
where
\beq{8.43}
\tilde J^{(1)}=\tilde x_0,\quad
\tilde J^{(2)}=\int\limits_0^{\tilde x_0}
\frac{\tilde Q_{\|}(\tilde Q_{\|}+2\tilde y)}
{\tilde x^2+(\tilde Q_{\|}+\tilde y)^2}\,
d\tilde x.
\eeq
Straightforward calculation gives
\beq{8.44}
\tilde J^{(2)}=\frac{2\tilde x_0}{3}.
\eeq

Then, summing the contributions from the regions (I) and (II) we get
in terms of the physical quantities the results
(\ref{8.45m})--(\ref{8.47m}).

\subsection{Inelastic transition probability for ${\bf G}\ne 0$}

In this part of Appendix we calculate the dimensionless factor
(\ref{8m.4}), which determines the inelastic transition probability
(\ref{8m.3})
with the momentum transfer ${\bf G}$ to the crystal lattice. The
final neutron energy $\ve'\simeq \hbar^2G^2/2m$ is much larger than
the barrier energy $U$ and initial energy $\ve$, thus
\beq{8m.1}
{\bf k}'_{\|}\simeq -{\bf G}_{\|},\quad
\bar k'\simeq k'_{\perp}\simeq G_{\perp}\gg\kp,\quad
r'\to 0,\quad
t'\to 1.
\eeq

The $\Pi$ factors (\ref{6m.30}) and (\ref{6m.31}) have now equal
contributions from backward and forward scattering and, with the help
of approximations (\ref{8m.1}) and expression
(\ref{6m.132})--(\ref{6m.136}), (\ref{5.16}), take the form
\beq{8m.8}
\Pi^{(1)}_l=
-\Frac{1}{\kp |z|^2}\,\,
{\rm Im}\left(
\Frac{\left({\bf G}_{\|}{\bf q}_{\|}+G_{\perp}\lambda_0\right)^2
U(\lambda_0,z)}
{q^2_{\|}+\lambda_0^2}
\right),
\eeq
\beq{8m.9}
\begin{array}{l}
\Pi^{(1)}_t=
-\Frac{G^2\,{\rm Im}\,U(\lambda_0,z)}{\kp |z|^2}+{}
\\[\bigskipamount]
\phantom{\Pi^{(1)}_t}+
\Frac{1}{\kp |z|^2}\,\,
{\rm Im}\left(
\Frac{\left({\bf G}_{\|}{\bf q}_{\|}+G_{\perp}\lambda_0\right)^2
U(\lambda_0,z)}
{q^2_{\|}+\lambda_0^2}
\right),
\end{array}
\eeq
\beq{8m.10}
\begin{array}{l}
\Pi^{(2)}_{l,t}=
\pm\Frac{q^2_{\|}|{\bf G}_{\|}{\bf q}_{\|}+G_{\perp}\lambda_0|^2-
|\lambda_0{\bf G}_{\|}{\bf q}_{\|}-G_{\perp}q^2_{\|}|^2}
{|q^2_{\|}+\lambda^2_0|^2}\times{}
\\[\bigskipamount]
\phantom{\Pi^{(2)}_{l,t}=
\pm q^2_{\|}|G_{\perp}\lambda_0-{\bf G}_{\|}{\bf q}_{\|}|^2}\times
\Frac{{\rm Im}\,U(iq_{\|},z)}{q_{\|}|z|^2},
\end{array}
\eeq
where
\beq{8m.7}
\lambda_0=i\kp+\xi_{\perp},\quad
\xi_{\perp}=k'_{\perp}-G_{\perp}.
\eeq

The integral (\ref{8m.4}) is over the Brillouin zone with the center
in ${\bf G}$. It is convenient to take
\beq{8m.11}
d^3k'=dk'_{\perp}d^2k'_{\|}\to d\xi_{\perp}d^2q_{\|},
\eeq
because ${\bf q}_{\|}=-{\bf k}'_{\|}-{\bf G}_{\|}$. Integration over
the angle between ${\bf q}_{\|}$ and ${\bf G}_{\|}$ gives
\beq{8m.12}
\av{{\bf G}_{\|}{\bf q}_{\|}}=0,\quad
\av{\left({\bf G}_{\|}{\bf q}_{\|}\right)^2\,}=
\frac{G^2_{\|}q^2_{\|}}{2}\,.
\eeq
Then the $W$ factors take the form
\beq{8m.13}
\begin{array}{l}
W^{(1)}_{l,t}(k_{\perp},k_{\|}\to {\bf G})=
\Frac{k_{\perp}}{\pi k\kp G^2}
{\ds\int} d\xi_{\perp}dq^2_{\|}\times{}
\\[\bigskipamount]
\phantom{}\times
\left(\mp\Frac{q^2_{\|}\left(G^2_{\|}\,/2-G^2_{\perp}\right)}
{|z|^2}\,
{\rm Im}\left(\Frac{U(\lambda_0,z)}{q^2_{\|}+\lambda^2_0}\right)-{}\right.
\\[\bigskipamount]
\left.\phantom{-q^2_{\|}(G^2_{\|}\,/2-G^2_{\perp})
(G^2_{\|}\,/2)}-
\Frac{G^2_{\perp,\|}\,{\rm Im}\,U(\lambda_0,z)}{|z|^2}\right),
\end{array}
\eeq
\beq{8m.14}
\begin{array}{l}
W^{(2)}_{l,t}(k_{\perp},k_{\|}\to {\bf G})=
\pm\Frac{k_{\perp}}{\pi kG^2}
{\ds\int} d\xi_{\perp}dq^2_{\|}\times{}
\\[\bigskipamount]
\phantom{}\times
\Frac{q_{\|}\left(G^2_{\|}\,/2-G^2_{\perp}\right)
(q^2_{\|}-|\lambda_0|^2)\,{\rm Im}\,U(iq_{\|},z)}
{|z|^2|q^2_{\|}+\lambda^2_0|^2}.
\end{array}
\eeq

Now we can integrate over $\xi_{\perp}$ transforming the integrals into
contour ones by closing a path in the complex $\xi_{\perp}$ plane. For
three functions to be integrated, one easily obtains
\beq{8m.15}
\int d\xi_{\perp}\,{\rm Im}\left(
\frac{U(\lambda_0,z)}{q^2_{\|}+\lambda^2_0}\right)=
\left\{\begin{array}{ll}
0, & q_{\|}<\kp,
\\[\smallskipamount]
\pi\, {\rm Im}\,U(iq_{\|},z)/q_{\|},\quad & q_{\|}>\kp,
\end{array}\right.
\eeq
\beq{8m.16}
\int d\xi_{\perp}\,{\rm Im}\,
U(\lambda_0,z)=
-2\pi x,
\eeq
\beq{8m.17}
\int d\xi_{\perp}
\frac{q^2_{\|}-|\lambda_0|^2}
{|q^2_{\|}+\lambda^2_0|^2}=
\left\{\begin{array}{ll}
-\pi/\kp,\quad & q_{\|}<\kp,
\\[\smallskipamount]
0, & q_{\|}>\kp.
\end{array}\right.
\eeq
Using these results we get for the sum of $W^{(1)}$ and $W^{(2)}$ factors
\beq{8m.18}
\begin{array}{l}
W_{l,t}(k_{\perp},k_{\|}\to{\bf G})=
\mp\Frac{k_{\perp}\left(G^2_{\|}\,/2-G^2_{\perp}\right)}
{k\kp G^2}\times{}
\\[\bigskipamount]
\phantom{{\cal T}_{l,t}(k_{\perp},k_{\|}\to{\bf G})={}}\times
{\ds\int} dq^2_{\|}
\Frac{q_{\|}\,{\rm Im}\,U(iq_{\|},z)}{|z|^2}+{}
\\[\bigskipamount]
\phantom{{\cal T}_{l,t}(k_{\perp},k_{\|})}+
\Frac{2k_{\perp} G^2_{\perp,\|}}{k\kp G^2}
{\ds\int} dq^2_{\|}\Frac{x}{|z|^2}.
\end{array}
\eeq

Using (\ref{6m.9.2}) and (\ref{8.39}), one obtains
\beq{8m.19}
\int dq^2_{\|}
\frac{q_{\|}\,{\rm Im}\,U(iq_{\|},z)}{|z|^2}=
-4J^{(2)},
\eeq
\beq{8m.20}
\int dq^2_{\|}\frac{x}{|z|^2}=2J^{(1)},
\eeq
where $J^{(1)}$ and $J^{(2)}$ are given by (\ref{8.44}). Thus, the
$W$ factors are of the form (\ref{8m.21m}) and (\ref{8m.21m.2}).


\begin{thebibliography}{99}

\bibitem{Lus69} V. I. Luschikov, Y. N. Pokotilovsky, A. V. Strelkov,
F. L. Shapiro, Sov. Phys. JETP Lett. {\bf 9}, 23 (1969).
\bibitem{Ste69} A. Steyerl, Phys. Lett. B, {\bf 29}, 33 (1969).
\bibitem{Gol96} R. Golub, Rev. Mod. Phys. {\bf 68} 329 (1996).
\bibitem{Kor04} E. Korobkina, R. Golub, J. Butterworth, P.
Geltenbort, S. Arzumanov, Phys. Rev. B {\bf 70} 035409 (2004).
\bibitem{Str00} A. V. Strelkov, V. V. Nesvizhevsky, P. Geltenbort,
D. G. Kartashov, A. G. Kharitonov, E. V. Lychagin, A. Yu. Muzychka,
J. M. Pendlebury, K. Schreckenbach, V. N. Shvetsov, A. P. Serebrov,
R. R. Taldaev, P. Yaidjiev, Nucl. Instr. Meth. A {\bf 440}, 695 (2000).
\bibitem{Bon02} L. N. Bondarenko, P. Geltenbort, E. I. Korobkina,
V. I. Morozov, Yu. N. Panin, Phys. At. Nucl. {\bf 65}, 11
(2002).
\bibitem{Lyc02} E. V. Lychagin, D. G. Kartashov, A. Yu. Muzychka, V. V. Nesvizhevsky, G. V. Nekhaev, A. V. Strelkov, Phys. At. Nucl. {\bf 65}, 1995 (2002).
\bibitem{Ste02} A. Steyerl, B. G. Yerozolimsky, A. P. Serebrov,
P. Geltenbort, N. Achiwa, Yu. N. Pokotilovski, O. Kwon,
M. S. Lasakov, I. A. Krasnoshchokova, A. V. Vasilyev, Eur. Phys. J. B
{\bf 28}, 299 (2002).
\bibitem{Ser03} A. P. Serebrov, J. Butterworth, M. Daum,
A. K. Fomin, P. Geltenbort, K. Kirch, I. A. Krasnoschekova,
M. S. Lasakov, Yu. P. Rudnev, V. E. Varlamov, A. V. Vassiljev,
Phys. Lett. A {\bf 309}, 218 (2003).
\bibitem{Lam02} S. K. Lamoreaux, R. Golub, Phys. Rev. C {\bf 66},
044309 (2002).
\bibitem{Hove54} L. Van Hove, Phys. Rev. {\bf 95}, 249 (1954).
\bibitem{Gur68} I. I. Gurevich, L. V. Tarasov, {\it Low-Energy
Neutron Physics} (Amsterdam, North Holland Pub. Co., 1968).
\bibitem{Lov84} S. W. Lovesey, {\it Theory of Neutron Scattering from
Condensed Matter}, vols. 1 and 2 (Oxford, Clarendon Press 1984)
\bibitem{Bar00} A. L. Barabanov, S. T. Belyaev, Eur. Phys. J. B
{\bf 15}, 59 (2000).
\bibitem{Ign90} V. K. Ignatovich, {\it The Physics of Ultracold
Neutrons} (Oxford, Clarendon Press, 1990).
\bibitem{Gol91} R. Golub, D. Richardson, S. K. Lamoreaux,
{\it Ultra-cold Neutrons} (Adam Hilger, Bristol, Philadelphia
and New York, IOP Publishing Ltd, 1991).
\bibitem{Blo77} D. I. Blokhintzev, N. M. Plakida, Phys. Status Solidi B {\bf 82}, 627 (1977); Preprint JINR P4-9631 (Dubna, JINR, 1976) (in Russian); Preprint JINR P4-10381 (Dubna, JINR, 1977) (in Russian).
\bibitem{Ign00} V. K. Ignatovich, M. Utsuro, Nucl. Instr. Methods A {\bf 440}, 709 (2000).
\bibitem{Ser05} A. Serebrov, N. Romanenko, O. Zherebtsov, M. Lasakov, A.
Vasiliev, A. Fomin, P. Geltenbort, I. Krasnoshekova, A.
Kharitonov, V. Varlamov, Phys. Lett. A {\bf 335}, 327 (2005).
\bibitem{Bar05} A. L. Barabanov, K. V. Protasov, nucl-th/0507020, Phys. Lett. A, in press.
\bibitem{Pok99a} Yu. N. Pokotilovski, Eur. Phys. J. B {\bf 8}, 1 (1999).
\bibitem{Pok99b} Yu. N. Pokotilovski, Phys. Lett. A {\bf 255}, 173 (1999).
\bibitem{Sch68} P. Schofield, In: {\it Physics of Simple Liquids},
Eds. H. N. V. Temperley, J. S. Rowlinson, G. S. Rushbrooke (North-Holland
Publishing Company, Amsterdam, 1968).
\bibitem{For75} D. Forster, {\it Hydrodynamic Fluctuations, Broken
Symmetry, and Correlation Functions} (W.A.Benjamin, Inc., Advanced
Book Program Reading, Massachusetts, London, Amsterdam, Don Mills,
Ontario, Sidney, Tokyo, 1975).
\bibitem{Akh81} A. I. Akhiezer, S. V. Peletminskii,
{\it Methods of statistical physics} (Pergamon Press, Oxford, New York, 1981).
\bibitem{Lan86} L. D. Landau, E. M. Lifshitz, A. P. Pitaevskii,
{\it Statistical physics} v.2 (Pergamon Press, Oxford, New York, 1986).
\bibitem{Lan86b} L. D. Landau, E. M. Lifshitz, A. M. Kosevich, L. P. Pitaevskii,
{\it Theory of elasticity} (Pergamon Press, Oxford, New York, 1986).
\bibitem{Bal69} B. Ya. Balagurov, V. G. Vaks, ZETPh {\bf 57}, 1646 (1969).
\bibitem{Fre46} Ia.I.Frenkel',
{\it Kinetic theory of liquids} (Oxford, Clarendon Press, 1946).
\bibitem{Nes02} V. V. Nesvizhevsky, Phys. At. Nucl. {\bf 65}, 400 (2002).

\end{thebibliography}
\end{document}